\documentclass[amsmath,amssymb,aps,prfluids]{revtex4}

\usepackage{graphicx}
\usepackage{subfig,enumerate}
\usepackage[mathscr]{euscript}
\usepackage{upgreek}
\usepackage{cancel}
\usepackage{dcolumn}
\usepackage{bm}
\usepackage{hyperref}
\usepackage{xcolor}

\begin{document}
\newcommand{\sm}[1]{{ #1}}
\newcommand{\nd}[1]{{ #1}}
\newcommand{\ub}{\mathbf{u}}
\newcommand{\eb}{\mathbf{e}}
\newcommand{\pard}[2]{\frac{\partial #1}{\partial #2}}
\newcommand{\Pe}{\mbox{Pe}}

\title{Instability and self-propulsion of active droplets along a wall}
\author{Nikhil Desai}
\author{S\'ebastien Michelin}%
 \email{sebastien.michelin@ladhyx.polytechnique.fr}
\affiliation{%
 LadHyX, D\'epartement de M\'ecanique, CNRS – Ecole Polytechnique, Institut Polytechnique de Paris, 91128 Palaiseau Cedex, France\\
}%

\date{\today}

\begin{abstract}
Active droplets can swim spontaneously in viscous flows as a result of the non-linear convective transport of a chemical solute produced at their surface by the Marangoni and/or phoretic flows generated by this solute's inhomogeneous distribution, provided the ratio of convective-to-diffusive solute transport, or P\'eclet number $\mbox{Pe}$ is large enough. As the result of their net buoyancy, active drops typically evolve at a small finite distance $d$ from rigid boundaries. Yet, existing models systematically focus on unbounded flows, ignoring the effect of the wall proximity on the intrisically-nonlinear nature of their propulsion mechanism. In contrast, we obtain here a critical insight on the propulsion of active drops near walls by analysing their stability to non-axisymmetric perturbations and the resulting emergence of self-propulsion along the wall with no limiting assumption on the wall-distance $d$. Dipolar or quadrupolar axisymmetric (levitating) base states are identified depending on $d$ and $\mbox{Pe}$. Perhaps counter-intuitively, a reduction in the drop-wall separation $d$ is observed to destabilize these modes and to promote  self-propulsion, as a result from the confinement-induced localisation of the chemical gradients driving the motion. In addition, quadrupolar states are more unstable than their dipolar counterparts due to the redistribution of the chemical perturbation by the base flow, favouring the emergence of stronger slip forcing on the drop surface.

\end{abstract}

\maketitle

\section{\label{s:intro}Introduction}

The study of synthetic micro-swimmers is one of the most rapidly advancing frontiers of modern science \cite{Maass2016, Moran2017, Wang2020}. The motion of these swimmers can be controlled and probed to a greater degree than that of more complex biological systems exhibiting similar behaviour \cite{Elgeti2015}. Experiments and analyses of the motion of artificial swimmers is therefore important for the comprehension of biological processes like planktonic patchiness \cite{Piontkovski1997, Abraham2000, Gouiller2021}, bio-convection \cite{Pedley1992, Bees2020, Kruger2016}, collective motion in bacterial colonies \cite{Sokolov2007, Theurkauff2012, Zottl2014}. From an application-oriented perspective, synthetic micro-swimmers are the foremost candidates toward accomplishing key tasks like payload manipulation and transport \cite{Burdick2008, Baraban2012}.

Synthetic micro-swimmers are generally inertialess, owing to the length-scales in which they operate, and must break time-reciprocal symmetries to swim in viscous flows~\cite{Purcell1977}. They can be divided into two broad categories, based on their propulsion strategies: (i) bio-mimetic structures that swim due to external forces \cite{Ghosh2009, Zhang2010, Ceylan2017}, and, (ii) chemically-powered entities that propel by extracting energy from their environment \cite{Maass2016, Moran2017, Wang2020}. The latter rely on two different physico-chemical properties, namely an \textit{activity}, $\mathcal{A}$, to produce a `fuel'/solute around them, and a \textit{mobility}, $\mathcal{M}$, to generate fluid forcing or fluid slip in response to inhomogeneous solute distributions~\cite{Anderson1989,Golestanian2007}. Canonical examples include phoretic particles that catalyze  chemical reactions on their surface~\cite{Paxton2004, Howse2007, Theurkauff2012, Ginot2018} and active drops that exchange surface-active agents with the surrounding fluid~\cite{Herminghaus2014, Izri2014, Maass2016}. Local inhomogeneities in solute concentration may then lead to a net slip velocity (diffusiophoresis) or net tangential fluid stress (Marangoni forcing) at the drop's surface.

A critical difference between Janus particles and active drops lies in the mechanism responsible for maintaining the solute's surface heterogeneity. Phoretic particles feature in general by design a chemical or geometric asymmetry~\cite{Golestanian2007, Michelin2015, Wang2020}, while chemically-active drops are intrinsically isotropic and break symmetry through an instability, that arises from the nonlinear convective transport of solute by the flow~\cite{Michelin2013,Izri2014}: small inhomogeneities in surface solute distribution generate fluid motion that further reinforces the chemical polarity of the drop at the origin of the flow. Convective solute transport therefore plays a key role in the dynamics of swimming droplets, i.e. the relevant P\'eclet number must be sufficiently large~\cite{Herminghaus2014, Izri2014, Maass2016}. In contrast, the smaller size of Janus swimmers and larger diffusivities of the involved solute, essentially result in a purely diffusive transport of solute species, i.e. $\Pe$ is close to zero~\cite{Moran2017}.

Depending on the nature of the droplet phase, suspending fluid and detailed surface activity, various swimming behaviors have been observed, e.g., rectilinear motion~\cite{Izri2014}, curling trajectories~\cite{Kruger2016b, Suga2018}, Brownian-motion-like diffusion \cite{Izzet2020}, chaotic oscillations~\cite{Suga2018}, mode-switching \cite{Hokmabad2021} and chemotaxis \cite{Jin2017}. Some of these behaviors have been analysed mathematically by considering the coupled dynamics of the convective and diffusive solute transport and that of the viscous flow, with model boundary conditions capturing the surface chemistry and interfacial dynamics. A key result has been the identification of a threshold ratio of advective-to-diffusive transport, above which isotropic rigid particles or drops undergo spontaneous diffusiophoretic and/or Marangoni propulsion~\cite{Yoshinaga2012, Michelin2013, Schmitt2013}. Such self-propulsion is consistent with the drag-reduction and reversal on heat-releasing drops~\cite{Rednikov1994b}. The relative strengths of phoretic and Marangoni effects do not affect the onset of propulsion~\cite{Morozov2019}, but do dictate the emergence of complex trajectories and swimming modes seen in experiments~\cite{Izzet2020, Hokmabad2021}. These swimming modes can also be recovered via more detailed descriptions of micellar solubilization (particularly relevant for supramicellar solutions), e.g., sorption of individual monomers onto the interface of a drop, followed by a release of filled micelles to the bulk~\cite{Morozov2020}. In addition to active drops in the bulk, symmetry-breaking through advection has been identified as a propulsion mechanism for otherwise symmetric \textit{interfacial} `boats': planar or partially immersed rigid particles that swim along an air-liquid interface~\cite{Boniface2019, Ender2021}. These propel via a surface-tension asymmetry along their periphery (line of contact on the air-liquid interface) and not via an induced slip velocity along the solid-liquid interface situated in the bulk. Yet, they display an important similarity: the requirement of strong enough advection for sustained propulsion. Collectively, the aforementioned studies demonstrate the robustness of the advective-coupling-induced swimming instability to a truly broad range of surface chemistry and physical arrangement.

Despite their ability to capture several key swimming behaviors, the studies on bulk motion of active drops share the common feature to analyse a single drop in an unbounded fluid domain, thus completely ignoring the presence of confining boundaries systematically observed in experiments~\cite{Kruger2016, Moerman2017, deBlois2019, Cheon2021} where active droplets swim along and close to rigid walls, due to the density difference with the suspending liquid. This underlines the pressing need to account for such confinement in the modeling of swimming drops, which is expected to  affect the solute transport around the drop (e.g. by preventing solute diffusion through the wall), the fluid motion and the droplet dynamics (e.g. modified mobility matrices).

Some of these features can already be observed in the much simpler limit of purely-diffusive transport ($\Pe=0$): the presence of a confining wall breaks the symmetry in the solute distribution around an isotropic particle resulting in the particle translation~\cite{Dominguez2016,Yariv2016, Yariv2016b}. In fact, $\Pe = 0$ yields a great simplification in terms of studying two-particle or particle-wall interactions, as convective solute transport can be safely ignored in that case, thus decoupling the chemical dynamics from the hydrodynamic flow. One can then solve in a first step the much simpler diffusion problem (Laplace equation) for the solute transport, and then proceed toward solving the fluid flow, which is simply a Stokes flow problem with boundary conditions that are known \textit{a priori}. This decoupling has indeed been exploited to study the general, non-axisymmetric interactions of phoretic Janus particles with a nearby wall~\cite{Mozaffari2016} and with each other~\cite{SharifiMood2016}.

 For active drops however, the convective transport can not be ignored as $\Pe>0$, and to obtain the drop's dynamics, the fully-coupled non-linear hydrochemical problem must be solved. Using a novel framework based on a body-fitted, deforming bi-spherical grid Lippera \textit{et al.} analysed axisymmetric (i.e. normal) collisions of a drop with a wall~\cite{Lippera2020}. Depending on $\Pe$, the wall-drop collisions were observed to be dominated by chemical interactions (moderate $\Pe$) or by a more complex hydrochemical coupling (larger $\Pe$). However, a complete understanding and modeling of the detailed hydrochemical coupling for droplet motions close to and along a wall still remains elusive, despite its experimental relevance~\cite{Kruger2016, Moerman2017, deBlois2019} and its potential impact on collective dynamics~\cite{Kruger2016,Thutupalli2018}.

This is precisely the focus of the present work, where we analyse the ability of an active drop to swim parallel and close to a rigid wall. We study the linear stability of the levitating non-propelling axisymmetric base state where the net wall-normal hydrodynamic force resulting from the solute polarity is balanced by the external force (e.g. gravity) maintaining the drop (which is intrinsically anti-chemotactic~\cite{Jin2017}) close to the wall. Indeed, completely force-free droplets or phoretic particles (i.e. with no external force) swim away from the wall in response to the accumulation of solute between the droplet/particle and the inactive wall. Here instead, the drop experiences a net external force $\mathbf{F}^\textrm{ext}$ normal to the wall and towards it, that balances the hydrodynamic force $\mathbf{F}_p$ and determines the equilibrium levitation distance $d^*$ of the base state. It is important to realize that this base state is completely generic: as long as there exists \textit{any} wall-normal, attractive external force on the drop, it will always ``find" the right hovering distance, $d^*$, from the wall. For example, a non-neutrally buoyant active drop will settle (or rise) toward the wall until it faces just enough phoretic repulsion to balance its (relative) weight, at which point it will cease to settle (or rise) and continue hovering at a fixed separation. In this way, the stationary nature of the drop is not artificially imposed via some well-tuned external force, but occurs naturally as dictated by the force balance. In this base state, the drop is not moving; but the wall-induced solute polarity drives a net fluid flow around the drop, which in turn modifies the chemical distribution, a fundamental difference with the unbounded configuration~\cite{Michelin2013}. Understanding the effect of such flows and resulting chemical distribution on the emergence of self-propulsion in the direction parallel to the wall is one of the key goals of this analysis. To this end, we expand the pertinent fields (flow and solute concentration) in terms of bispherical coordinate eigenfunctions, and analyse the stability of the levitating base state with respect to non-axisymmetric perturbations. We thus determine the conditions for self-propulsion in the horizontal (wall-parallel) direction, in which the droplet experiences no external forces.

The rest of the paper is organized as follows. Section~\ref{s:mathMod} introduces the physical problem in more details along with the relevant governing equations for the coupled dynamics of the solute concentration and flow fields.  In Section~\ref{s:baseState}, we determine the axisymmetric steady solutions of these coupled problem and describe in detail the key features of the possible levitating base state as a function of $\Pe$ and drop-wall distance $d$. Section~\ref{s:stability} then introduces the relevant linearised equations for the non-axisymmetric perturbations and the solution methodology for the linear stability analysis, whose results are analysed in Section~\ref{s:results}. We finally summarise our main findings in Section~\ref{s:conclusion} and present some further perspectives on future work.

\section{Mathematical model} \label{s:mathMod}

\subsection{Problem description} \label{s:mathMod1}

\begin{figure}[t]
\begin{center}
      \includegraphics[width=6cm]{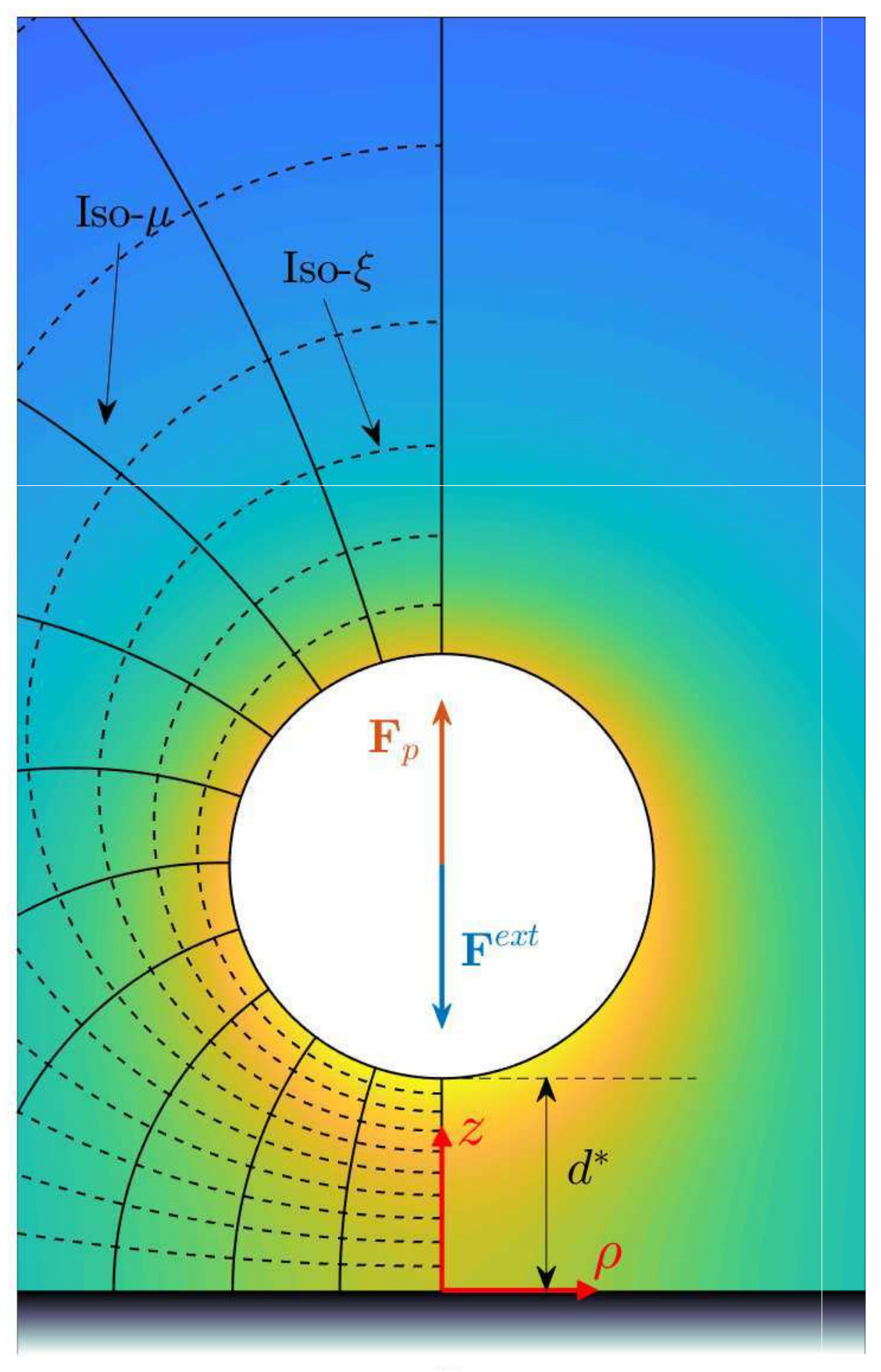}
  \caption{An active drop levitating above a rigid wall. The equilibrium distance, $d^*$, is dictated by a balance between the external force, $\textbf{F}^{ext}$, and the `phoretic force', $\textbf{F}_p$. The iso-surfaces of constant $\xi$ and $\mu$ on the left half, define the (body-fixed) bi-spherical grid. Also shown is the origin of the $\left( \rho, z \right)$ cylindrical coordinate system.}
  \label{probSchem}
\end{center}
\end{figure}
We focus on the near-wall dynamics of an active droplet of radius $R$, that is initially levitating in equilibrium at a distance $d^*$ from a rigid, passive wall. Its slow solubilisation in the surrounding fluid is modelled here as the emission of a solute at a rate $\mathcal{A}>0$ (e.g. emission of filled micelles). The bulk concentration of this solute is $c^*$ and its molecular diffusivity is $D$. The interplay between phoretic and Marangoni effects on the surface of the drop may be quite complex, in particular for surfactant-loaded interfaces. So we focus on a purely phoretic approach for simplicity, such that an inhomogeneity in solute concentration around the droplet causes a phoretic slip on the drop's surface, characterized by a mobility $\mathcal{M}>0$. This fluid slip can lead to the swimming motion of the inertia-less drop. It also drives flow in the suspending fluid, which redistributes the emitted solute. The fluid flow and solute transport are thus strongly coupled, and one must \textit{simultaneously} solve for the flow field and the solute concentration around the drop to characterise the droplet's self-propulsion.

\subsection{Governing equations and boundary conditions} \label{s:mathMod2}

We are interested to study the onset of motion of an active drop parallel to a confining wall. The physics of this problem is governed by the advection-diffusion equation for the solute concentration and the continuity and Stokes equations for fluid flow. The latter is forced by a slip velocity on the surface of the drop, $\textbf{u}^*_s = \mathcal{M} \nabla^*_s c^*$ (asterisks denote dimensional quantities), thus coupling the fluid flow and the solute transport. Defining the characteristic phoretic velocity $V_c=\mathcal{AM}/D$, all variables are non-dimensionalised in the following using $R$, $V_c$, $R/V_c$, $\eta V_c/R$ and $R\mathcal{A}/D$ as characteristic length, velocity, time, pressure and concentration, with $\eta$ the outer fluid's viscosity. The dimensionless advection-diffusion equation thus becomes:
\begin{equation}\label{AdvDiff_gde}
    \frac{\partial c}{\partial t} + \textbf{u} \cdot \nabla c = \frac{1}{Pe} \nabla^2 c,
\end{equation}
where $\Pe=V_cR/D$ is the P\'eclet number. Eq.~\eqref{AdvDiff_gde} is complemented with the boundary conditions,
\begin{equation}\label{AdvDiff_bcs}
    {\left. {\textbf{n} \cdot \nabla c} \right|}_{\mathscr{W}} = 0, \qquad{\left. {\textbf{n} \cdot \nabla c} \right|}_{\mathscr{S}} = -1,
\end{equation}
where $\textbf{n}$ is the outward pointing normal on the surface of the drop, and $\mathscr{W}$ and $\mathscr{S}$ denote the wall and the drop surface, respectively. In addition, the concentration vanishes far away from the drop,
\begin{equation}\label{AdvDiff_bcs_inf}
    {\left. c \right|}_{r \to \infty} = 0.
\end{equation}
In Eq.~\eqref{AdvDiff_gde}, $\textbf{u}$ is the velocity field in the fluid surrounding the drop. The fluid flow is governed by the incompressible Stokes equations:
\begin{equation}\label{continuity}
    \nabla \cdot \textbf{u} = 0, \qquad \nabla p=\nabla^2 \mathbf{u}.
\end{equation}
with the velocity vanishing at the wall,
\begin{equation}\label{Stokes_bcs_wall}
    {\left. \textbf{u} \right|}_{\mathscr{W}} = \textbf{0}.
\end{equation}
The fluid velocity vanishes away from the drop (fluid at rest)
\begin{equation}\label{Stokes_bcs_inf}
    {\left. \textbf{u} \right|}_{r \to \infty} = \textbf{0},
    \end{equation}
and is forced at the drop surface by the local, surface gradient of the solute concentration, and the drop's translation and rotation,
\begin{equation}\label{Stokes_bcs_drop}
    {\left. \textbf{u} \right|}_{\mathscr{S}} = \nabla_s c + \textbf{V} + \textbf{W} \times \textbf{x}_s,
\end{equation}
where, $\textbf{x}_s$ is the position vector from the center of the drop to its surface, and $\textbf{V}$ and $\textbf{W}$ are respectively the translational and rotational velocities.

The latter are obtained by enforcing that the drop must experience zero total force (resp. torque) at all times. In the presence of an external force $\textbf{F}^{ext}$ and no external torque, these conditions are given by:
\begin{equation}
\label{force_int_gen}
    \int_{S} { \textbf{n} \cdot \boldsymbol\sigma dS } + \textbf{F}^{ext} = \textbf{0},\qquad  \int_{S} { \textbf{x}_s \times \left( \textbf{n} \cdot \boldsymbol\sigma \right)dS } = \textbf{0},
\end{equation}
where $\boldsymbol\sigma$ is the stress tensor in the fluid.

\section{The non-quiescent base state} \label{s:baseState}

In this section, Eqs.~\eqref{AdvDiff_gde} to \eqref{force_int_gen} are solved to identify the influence of the wall on the fluid flow and solute concentration around the drop to identify steady axisymmetric equilibrium or `hovering' states of the drop ($\mathbf{V}=\mathbf{W}=\textbf{0}$). We re-emphasize here that in an experimental realization, the hovering states can be naturally obtained without the need to tune the external force experienced by the drop. Such base states are necessarily non-quiescent: the inherent asymmetry in solute distribution will cause a slip on the drop surface, which will then drive bulk fluid flow.

\subsection{Solution methodology: Axisymmetric bi-spherical coordinates} \label{s:baseState1}
The motion of a spherical particle near a wall (or, another sphere) is quintessential for the utilization of a bi-spherical coordinate system, as demonstrated in countless phoretic interaction problems \cite{Popescu2011, Michelin2015, SharifiMood2016, Mozaffari2016, Nasouri2020, Nasouri2020b}. The cylindrical coordinates $\left( \rho, z, \phi \right)$, made dimensionless using the scales specified in Section \ref{s:mathMod2}, are expressed in the bi-spherical coordinates $\left( \xi, \mu, \phi \right)$ as:
\begin{equation}\label{rho_z_xi_mu}
    \rho = \frac{a \sqrt{1 - \mu^2}}{\Gamma(\xi,\mu)}, \;
    z = \frac{a \sinh(\lambda \xi)}{\Gamma(\xi,\mu)},
\end{equation}
where, $\Gamma(\xi, \mu) = \left( \cosh(\lambda \xi) - \mu \right)$, $\lambda = \cosh^{-1}(1+d)$ and $a = \sqrt{d(d+2)}$; with $d$ being the dimensionless drop-wall separation, i.e., $d=d^*/R$. The coordinate system is defined such that $z=0$ denotes the wall and the $z$-axis passes through the center of the drop. In the bi-spherical coordinate system, $\xi = 0$ corresponds to the wall and $\xi = 1$ to the surface of the drop (see Fig. \ref{probSchem}). The basis vectors in the bi-spherical coordinates, $\left( \textbf{e}_{\xi}, \textbf{e}_{\mu} \right)$, are related to those in the cylindrical coordinates, $\left( \textbf{e}_{z}, \textbf{e}_{\rho} \right)$, as:
\begin{align}\label{erho_ez_exi_emu}
    \textbf{e}_{\xi} = \frac{1-\mu \cosh (\lambda \xi)}{\Gamma(\xi,\mu)}\textbf{e}_{z} - \frac{\sqrt{1-\mu^2} \sinh(\lambda \xi)}{\Gamma(\xi,\mu)}\textbf{e}_{\rho}, \nonumber \\
    \textbf{e}_{\mu} = \frac{\sqrt{1-\mu^2} \sinh(\lambda \xi)}{\Gamma(\xi,\mu)}\textbf{e}_{z} + \frac{1-\mu \cosh (\lambda \xi)}{\Gamma(\xi,\mu)}\textbf{e}_{\rho}.
\end{align}

\subsubsection{The hydrodynamics problem: streamfunction formulation} \label{s:baseState1a}

Axisymmetric solutions of the Stokes equation for the  flow are conveniently described in terms of a  streamfunction, defined in bi-spherical coordinates as

\begin{equation}\label{u}
    \ub=\frac{\Gamma^2}{a^2}\left(\pard{\psi}{\mu}\eb_\xi-\frac{1}{\lambda\sqrt{1-\mu^2}}\pard{\psi}{\xi}\eb_\mu\right),
\end{equation}
where
\begin{equation}\label{strFun_def}
    \psi\left( \xi, \mu, t \right) = \Gamma^{-3/2}\sum\limits_{n=0}^{\infty }{\left( 1-{{\mu }^{2}} \right){{{{L}'}}_{n}}\left( \mu  \right){{U}_{n}}\left( \xi, t  \right)},
\end{equation}
and
\begin{align}\label{Un}
    U_n\left( \xi, t \right) & = \alpha_n(t) \cosh \left\{ (n+3/2) \lambda \xi \right\} + \beta_n(t) \sinh \left\{ (n+3/2) \lambda \xi \right\} \nonumber \\ & + \gamma_n(t) \cosh \left\{ (n-1/2) \lambda \xi \right\} + \delta_n(t) \sinh \left\{ (n-1/2) \lambda \xi \right\}.
\end{align}
Here, $L_n(\mu)$ is the Legendre polynomial of degree $n$, and prime denotes the derivative of a one-variable function. The flow field is completely specified by the knowledge of the constants $\left[ \alpha_n, \beta_n, \gamma_n, \delta_n \right]$, that are determined from boundary conditions \eqref{Stokes_bcs_wall} and \eqref{Stokes_bcs_drop} at each instant $t$. The impermeability of the wall and drop imposes $\psi$ to take constant values along each. Furthermore, the problem being axisymmetric, \nd{$\rho=0$} is also a streamline, so that 
\begin{equation}\label{u_norm_wall_drop}
    \left. \psi \right|_{\xi = 0} = \left. \psi \right|_{\xi = 1} = 0.
\end{equation}
The tangential velocity boundary condition at the wall is simply,
\begin{equation}\label{u_tan_wall}
    \left. \frac{\partial \psi}{\partial \xi} \right|_{\xi = 0} = 0.
\end{equation}
Finally, the phoretic slip boundary condition at the surface of the stationary drop yields,
\begin{equation}\label{u_tan_drop}
    \left.\frac{\partial \psi}{\partial \xi} \right|_{\xi = 1} = \left. -\frac{a\lambda(1-\mu^2)}{\Gamma} \frac{\partial c}{\partial \mu} \right|_{\xi = 1}.
\end{equation}
Truncation of the expansion of Eq.~\eqref{strFun_def} at a finite $N$ number of terms and projection of Eqs.~\eqref{u_norm_wall_drop} to \eqref{u_tan_drop} onto the $n$-th Legendre polynomial $L_n \left( \mu \right)$, provides four sets of $N$ linear equations each, relating the coefficients $\left[ \alpha_1 ... \alpha_N, \beta_1 ... \beta_N, \gamma_1 ... \gamma_N, \delta_1 ... \delta_N \right]$ to the solute's surface concentration distribution (via Eq.~\eqref{u_tan_drop}). \nd{This makes the hydrodynamics problem linear as well as instantaneous in the solute concentration field, $c\left( \xi, \mu, t \right)$. The solution of the solute concentration is summarized next.}

\subsubsection{The solute transport problem} \label{s:baseState1b}

The solute concentration can be decomposed in terms of azimuthal components, as
\begin{equation}\label{c_axisymm}
    c\left( \xi, \mu, t  \right) = \Gamma^{1/2}\sum\limits_{n=0}^{\infty }{{{{{L}}}_{n}}\left( \mu  \right){{c}_{n}}\left( \xi, t  \right)}.
\end{equation}
Since the flow-field is fully determined once the surface concentration is known, the base state solution essentially requires us to determine steady solutions of $c_n(\xi)$. To this end, the flow field solution obtained in the previous subsection is substituted in the advection-diffusion equation, Eq.~\eqref{AdvDiff_gde}. After substitution of Eqs.~\eqref{u} and \eqref{c_axisymm} into Eq.~\eqref{AdvDiff_gde}, the successive projections of Eq.~\eqref{AdvDiff_gde} onto $L_p \left( \mu \right)$ are rewritten formally as
\begin{equation}\label{AdvDiff_axiSymm_disc}
    \mathbf{H}^1 \cdot \frac{\partial \mathbf{C}}{\partial t} + \left( {{\mathbf{B}}^{1}}\cdot \mathbf{U}+{{\mathbf{B}}^{2}}\cdot \frac{\partial \mathbf{U}}{\partial \xi } \right)\cdot \mathbf{C}+\left( {{\mathbf{B}}^{3}}\cdot \mathbf{U} \right)\cdot \frac{\partial \mathbf{C}}{\partial \xi }=\frac{1}{Pe}\left\{ {{\mathbf{A}}^{1}}\cdot \mathbf{C}+{{\mathbf{A}}^{2}}\cdot \frac{{\partial ^{2}}\mathbf{C}}{\partial {{\xi }^{2}}} \right\},
\end{equation}
where, $\textbf{C} \equiv \left[ c_0(\xi, t),\;c_1(\xi, t),\;...,\;c_N(\xi, t) \right]^T$ is the set of unknown concentration components and $\textbf{U} \equiv \left[ U_1(\xi, t),\;U_2(\xi, t),\;...,\;U_N(\xi, t) \right]^T$ is the vector defining the streamfunction (Eq.~\eqref{strFun_def}), which is known at each instant as a linear function of $\mathbf{C}$. The third order tensors $\textbf{B}^i$ and the second order tensors $\textbf{H}^1$, $\textbf{A}^i$ depend only on $\xi$ (see Ref.~\cite{Lippera2020} for more details, and their complete expressions in the Appendix). The applicable boundary conditions for these second-order equations in $\xi$ are obtained by projecting  Eq.~\eqref{AdvDiff_bcs} onto \nd{$L_n(\mu)$ (with $n \ge 0$):
\begin{equation}\label{AdvDiff_axiSymm_disc_BCs1}
    \left. \left( \frac{\partial c_n}{\partial \xi} - \frac{n+1}{2n+3} \frac{\partial c_{n+1}}{\partial \xi} -\frac{n}{2n-1} \frac{\partial c_{n-1}}{\partial \xi} \right) \right|_{\xi=0} = 0,
\end{equation}
and,
\begin{equation}\label{AdvDiff_axiSymm_disc_BCs2}
    \left. \left( \frac{\lambda \sinh(\lambda)}{2}c_n + \cosh(\lambda) \frac{\partial c_n}{\partial \xi} - \frac{n+1}{2n+3} \frac{\partial c_{n+1}}{\partial \xi} - \frac{n}{2n-1} \frac{\partial c_{n-1}}{\partial \xi} \right) \right|_{\xi=1} = \sqrt{2}a \lambda \textrm{e}^{ -\lambda(n+1/2) }.
\end{equation}}

\subsubsection{Numerical solution}

The vector $\mathbf{U}$ is obtained directly, at each instant, from the boundary values $c_n(\xi=1)$ of the concentration components. In the following, we solve Eqs.~\eqref{AdvDiff_axiSymm_disc} to \eqref{AdvDiff_axiSymm_disc_BCs2} numerically  by discretizing $\xi$-derivatives using second-order centered finite differences over a uniform grid of $M$ equispaced points between $\xi=0,1$. An explicit time-marching is used for the advective term while diffusive terms are treated using the Crank-Nicholson scheme. We employ a continuation procedure to obtain the steady state solutions for different values of the P\'eclet number $\mbox{Pe}$, starting from the solution for $\Pe=0$, and increasing $\mbox{Pe}$ incrementally: at a given $\mbox{Pe}_k$, the initial condition for $\textbf{C}$ is chosen as the steady-state solution at the previous increment $\mbox{Pe}_{k-1}$. After a steady-state solution is obtained, the vertical (hydrodynamic) force experienced by the drop, $F_z$, can be evaluated in terms of the velocity coefficients $\left[ \alpha_n, \beta_n, \gamma_n, \delta_n \right]$ as:
\begin{equation}\label{Fz_axisymm_num}
    F_z = -\frac{2\sqrt{2}\pi}{a}\sum\limits_{n=1}^{\infty }{n\left( n+1 \right) \left( \alpha_n + \beta_n + \gamma_n + \delta_n \right)},
\end{equation}
where the force has been made dimensionless by the quantity $\eta V_c R$. The force $\textbf{F}_p$ in Fig. \ref{probSchem} is thus, $\textbf{F}_p = F_z\textbf{e}_z$.

\subsubsection{Validation of the base state solution} \label{s:baseState1c}

No net force is obtained when the wall-to-droplet distance is infinite. For large separations, i.e. $d\gg 1$, the force can be obtained  asymptotically as (see Appendix)
\begin{equation}\label{Fz_axisymm_ana}
    F_z^{ana} = -\frac{12 \pi}{\left( d+1 \right)^2 \left( Pe-8 \right)}\cdot
\end{equation} 
The expression in Eq.~\eqref{Fz_axisymm_ana} is singular for $\Pe = 8$. This should be no surprise: in the unbounded case ($d \to \infty$), the non-linear advective coupling in the solute transport causes a second instability of the isotropic solution for $\Pe=8$, corresponding to a spontaneous pumping mode where the particle is fixed and experiences a net force. The asymmetry introduced by the wall acts as a forcing of this instability which resonates at $\Pe=8$ in the large-distance limit. \nd{It is crucial to note here that this `pumping instability' is distinct from the `swimming instability' studied in Ref.~\cite{Michelin2013}, which is triggered for $\Pe \ge 4$. This is the result of the drop's being maintained at a fixed height along $\eb_z$ here, a constraint that is absent in Ref.~\cite{Michelin2013}. If the drop is kept fixed, then dipolar deviations from the isotropic solute distribution result in dipolar flows that decay as $\sim 1/r_d$ away from the drop ($r_d$ being the radial coordinate measured from the drop center), as opposed to the $\sim 1/r_d^3$ decay for a non-fixed (force-free) drop. This fundamentally alters the nature of instability of the first azimuthal concentration mode ($n=1$), which now sets in at $\Pe=8$ (as a pumping mode), instead of the swimming mode at $\Pe=4$ in Ref.~\cite{Michelin2013}. Note also that \textit{only} the instability threshold associated with the $n=1$ concentration mode is altered for a fixed drop, while those corresponding to the higher azimuthal modes remain unchanged. Again, this is because the structure of the higher order flow-fields remains the same for a fixed and a non-fixed drop. An intriguing consequence of the different critical P\'eclet numbers for the swimming and pumping instabilities is the existence of a range, $4 \le \Pe \le 8$, where we obtain force-free and moving solutions (i.e., $\textbf{F} = \textbf{0}$ and $\textbf{U} \ne \textbf{0}$) but \textit{do not} obtain forced and non-moving solutions (i.e., $\textbf{F} \ne \textbf{0}$ and $\textbf{U} = \textbf{0}$). This is in stark contrast with conventional swimmers, which always require a non-zero external force in order to completely arrest their motion. Such non-trivial dynamics is also observed in heat-releasing or surfactant-consuming drops~\cite{Rednikov1994b, Rednikov1994}, where the pumping instability can occur at a higher critical Marangoni number than the swimming instability.}

\begin{figure}
\begin{center}
      \includegraphics[width=8cm]{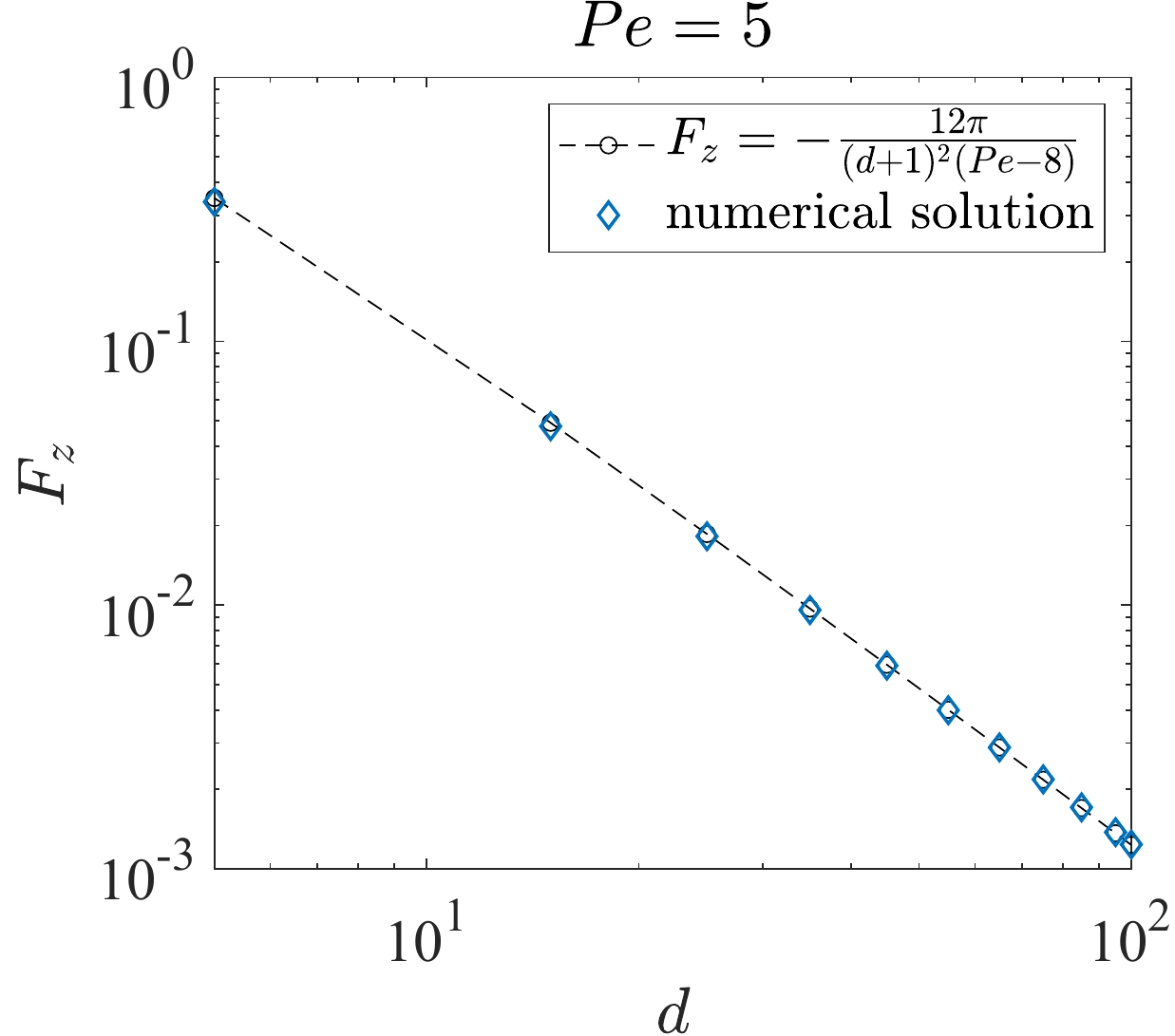}
  \caption{The vertical force acting on the (fixed) drop as a function of its separation, $d$, from the wall. The symbols represent results from the numerical analysis described in Section \ref{s:baseState} and the dashed line is the analytical approximation for the case when $d >> 1$ and $\Pe < 8$.}
  \label{Fz_vs_d_valid}
\end{center}
\end{figure}

We validate our numerical results by comparing the force obtained numerically (Eq.~\eqref{Fz_axisymm_num}) to the analytical solution in the limit $d \gg 1$, Eq.~\eqref{Fz_axisymm_ana}. Fig. \ref{Fz_vs_d_valid} shows an excellent match between the two solutions, with the numerical solution obtained for $N=60$ azimuthal modes and $M=200$ uniformly spaced points between $0 \le \xi \le 1$. We now use our numerical method to study the dependence of the base state concentration and flow-fields on the physical parameters of interest: the drop-wall separation, $d$, and the P\'eclet number, $\Pe$.

\subsection{Characteristics of the axisymmetric base state} \label{s:baseState2}

\begin{figure}[t!]
 \begin{center}
 \includegraphics[width=14cm]{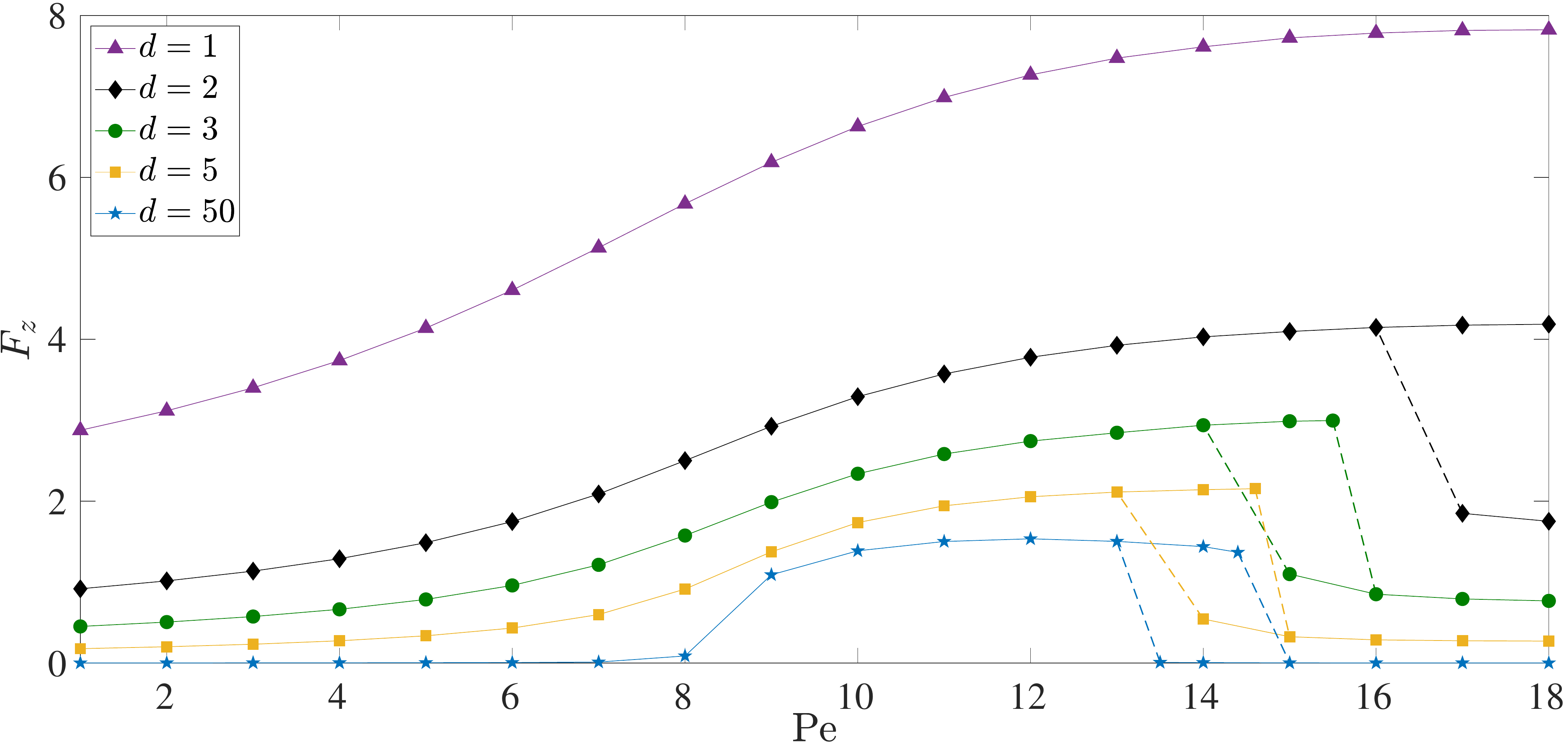}\\
 \vspace{.5cm}
    \includegraphics[width=4.75cm]{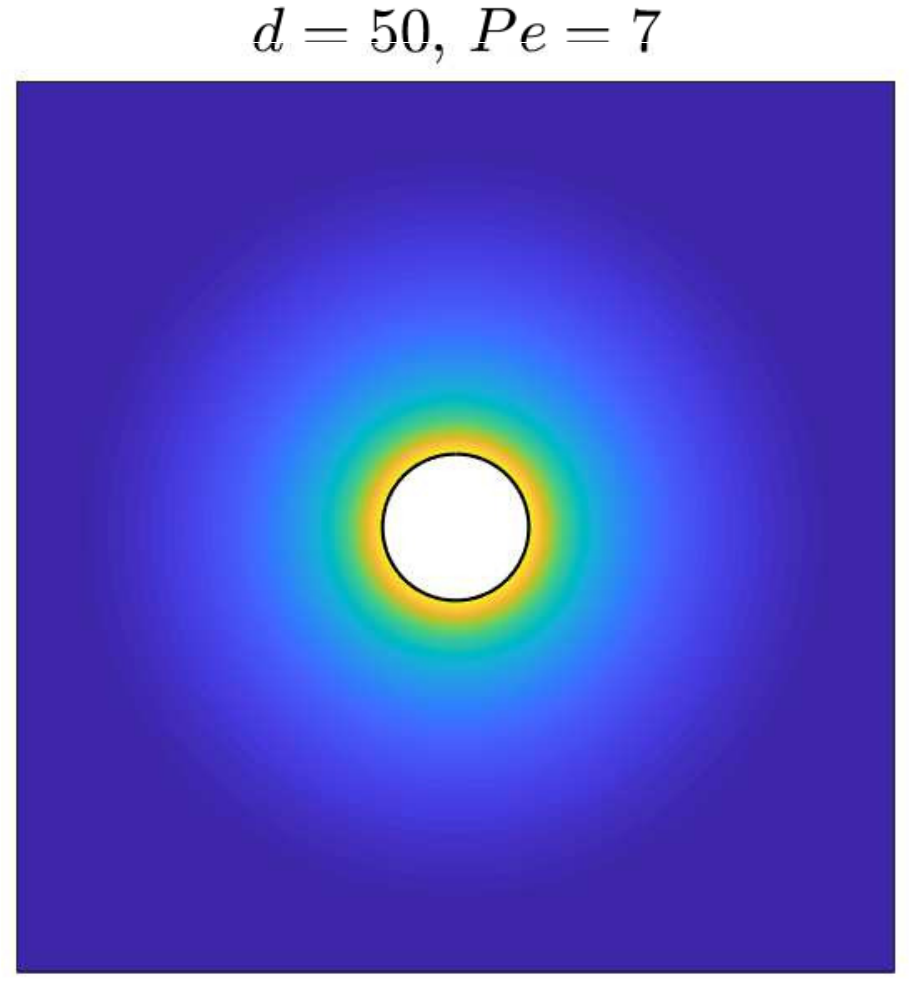} \hspace{0.1cm}
      \includegraphics[width=4.75cm]{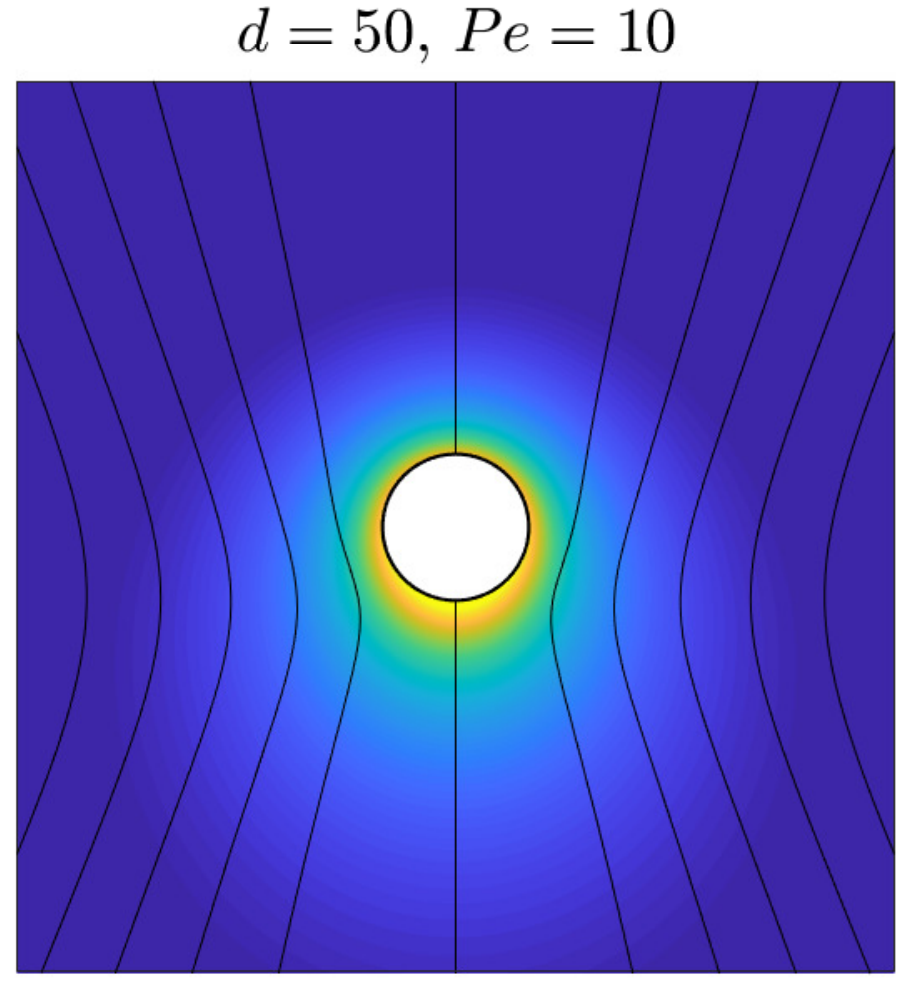} \hspace{0.1cm}
      \includegraphics[width=4.75cm]{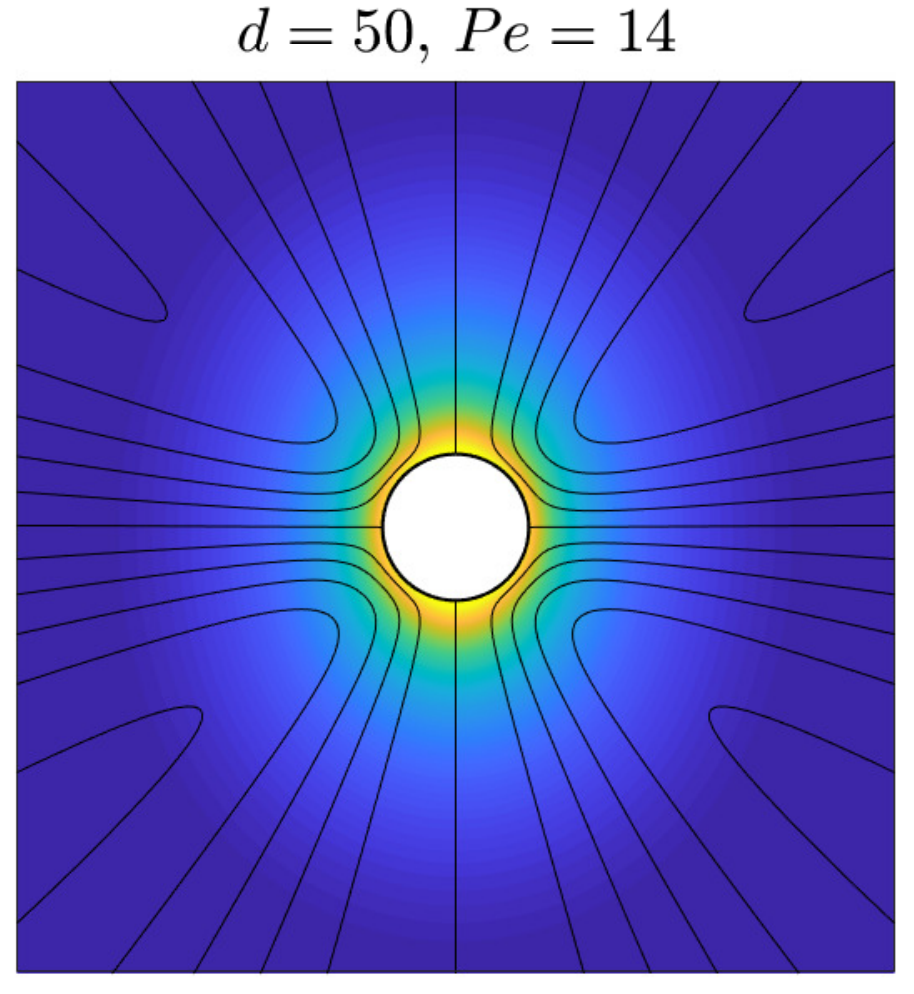}\\
      \vspace{0.2cm}
      \includegraphics[width=4.75cm]{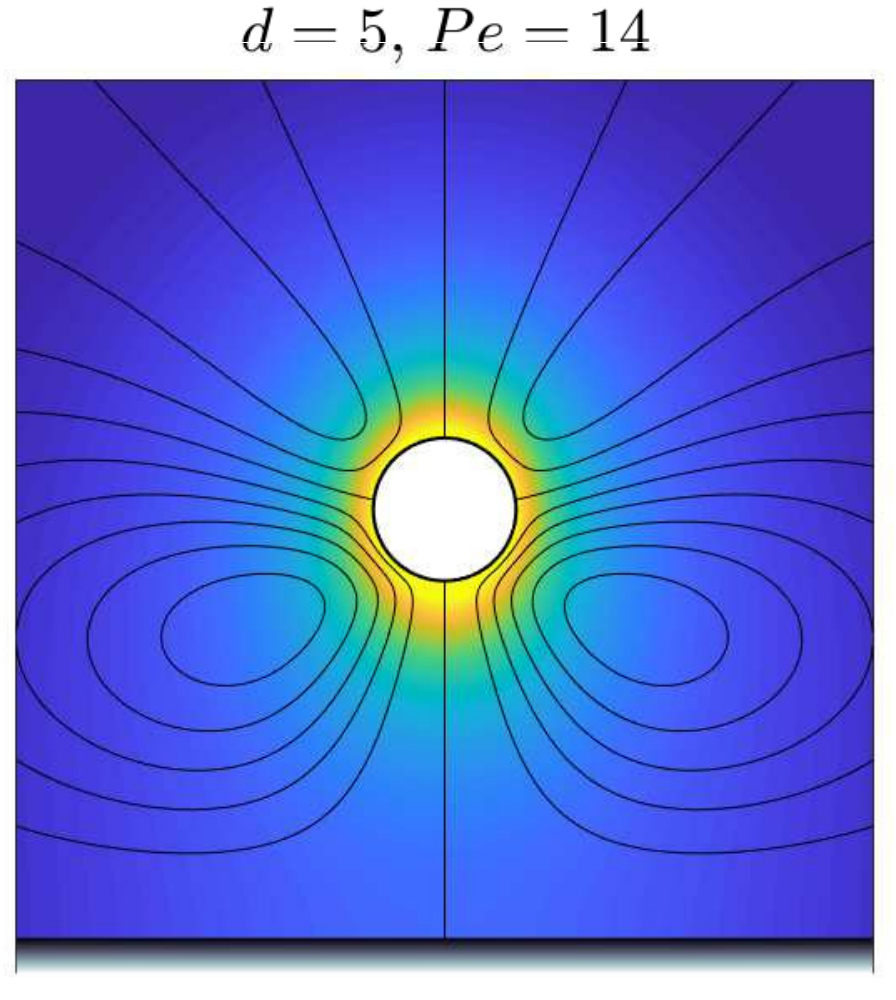} \hspace{0.1cm}
      \includegraphics[width=4.75cm]{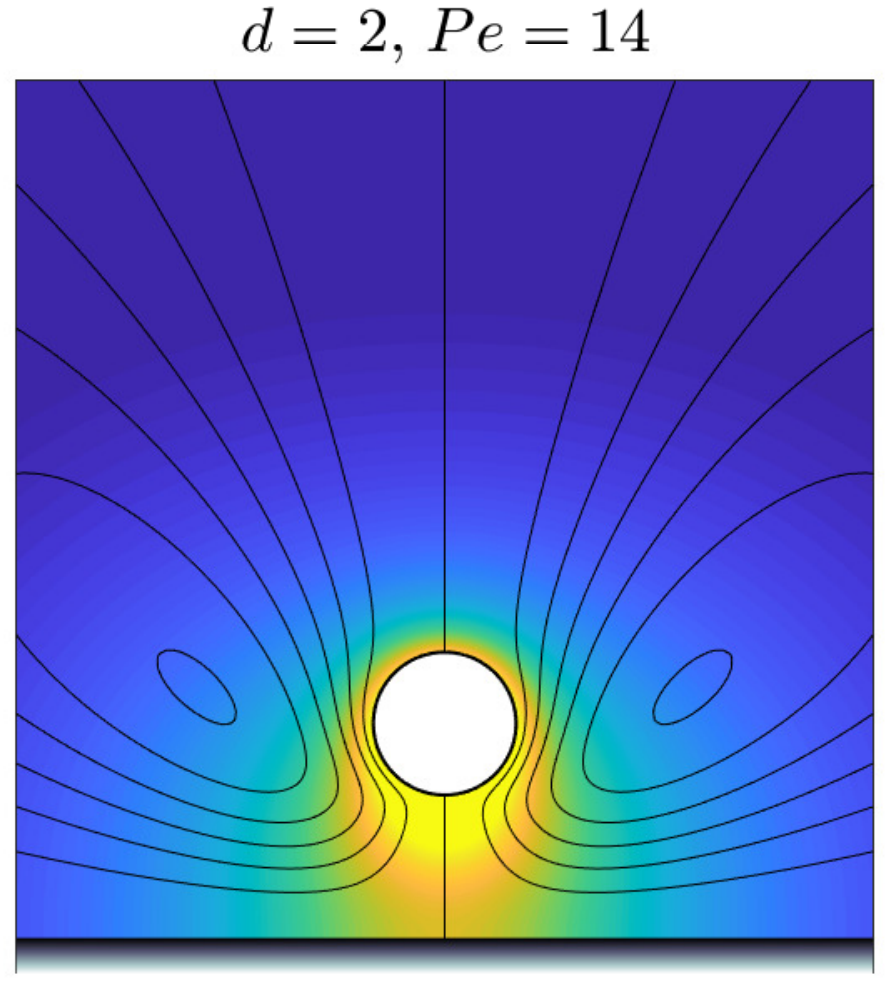} \hspace{0.1cm}
      \includegraphics[width=4.75cm]{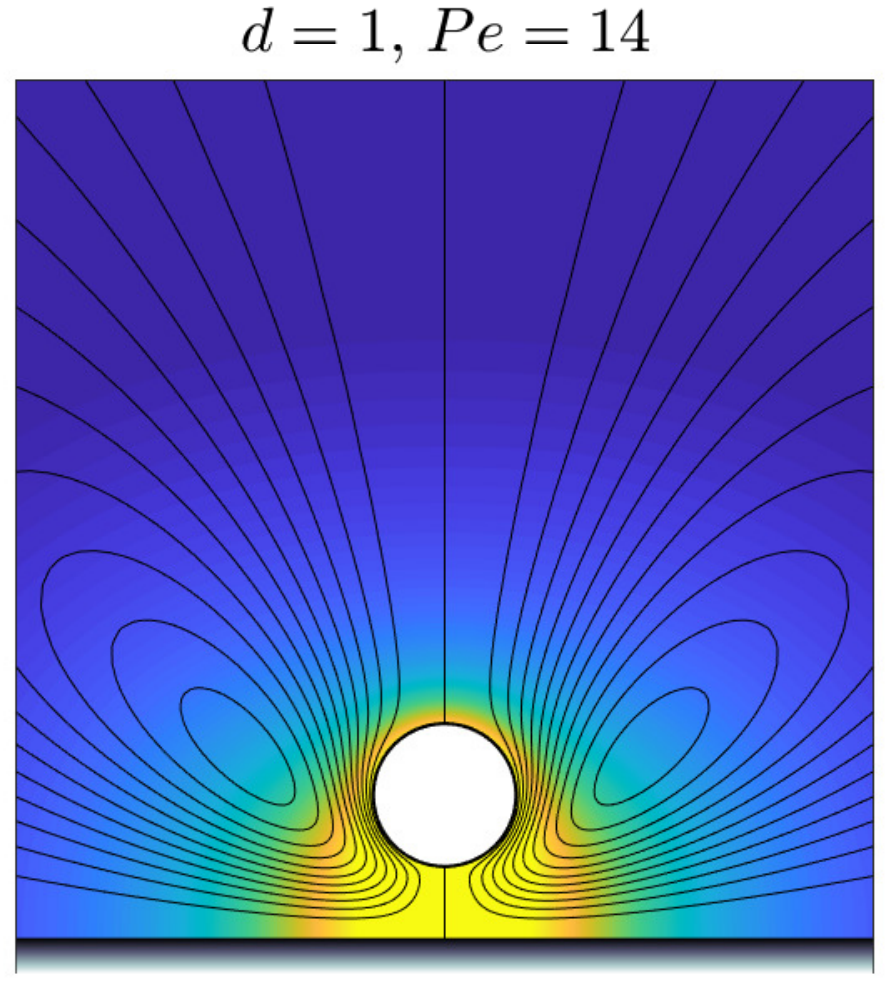}
  \caption{Characteristics of the axisymmetric base state. (Top) Vertical force, $F_z$, on the drop as a function of the P\'eclet number, $\Pe$, for different drop-wall separation distance, $d$. (Center) Solute distribution and streamlines (iso-lines of the streamfunction, $\psi$) around the drop for $d=50$ (representative of the unbounded limit $d\rightarrow \infty$) for $\Pe=7$ (left), $\Pe=10$ (center) and $\Pe=14$ (right). No streamlines are shown for $\Pe=7$, as the fluid is essentially at rest. (Bottom) Solute distribution and streamlines for $\Pe=14$ for $d=5$ (left), $d=2$ (center) and $d=1$ (right).}
  \label{Fz_vs_Pe_d}
 \end{center}
\end{figure}

\begin{figure}[t]
    \begin{center}
    \includegraphics[width=5.5cm]{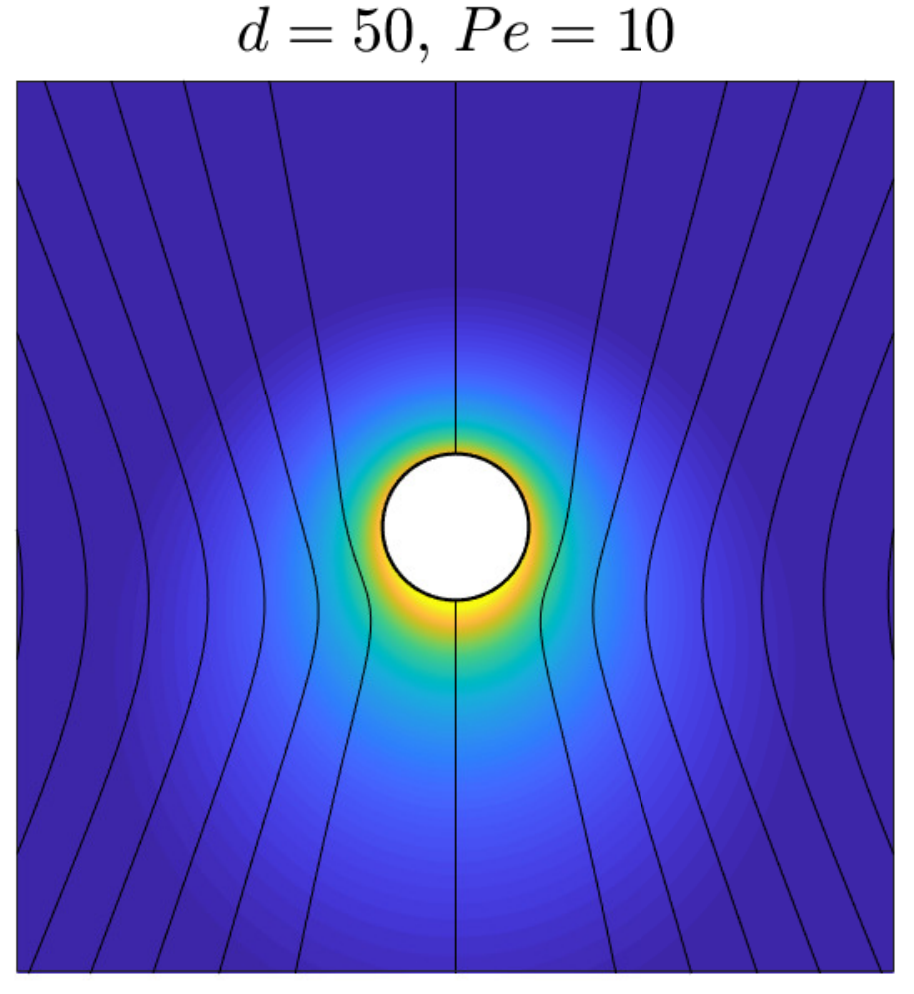}
    \hspace{1cm}
    \includegraphics[width=5.5cm]{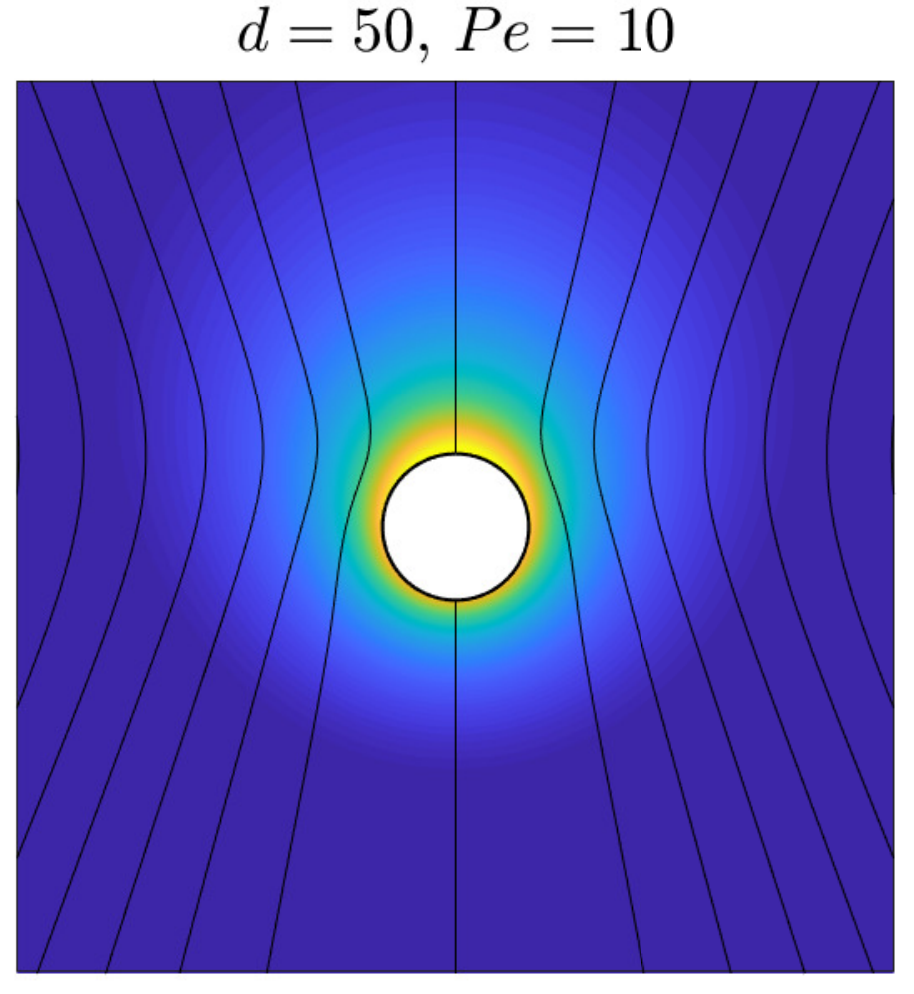}\\
    \caption{The steady-state solution branches of the symmetric bifurcation observed beyond $\Pe=8$ in the base state for $d \to \infty$, as shown in Fig. \ref{Fz_vs_Pe_d} and discussed in Section \ref{s:baseState}. $F_z > 0$ (resp. $F_z < 0$) for the solution on the left (resp. right).}
    \label{symm_bif}
    \end{center}
    \end{figure}

We first discuss the nature of the base state in the unbounded limit, i.e., for $d \to \infty$. 
For $\Pe<8$ and $d \to \infty$, the solute distribution around the drop is essentially isotropic; there is therefore no fluid flow and the drop remains force-free, i.e., $F_z = 0$ (Figure~\ref{Fz_vs_Pe_d}, left). For $\Pe \ge 8$ and $d \to \infty$, the system exhibits a symmetric bifurcation and we observe a non-quiescent steady-state, i.e., one where the solute concentration along the drop’s surface is non-uniform, leading to an induced slip velocity on the drop’s surface, which then drives fluid flow around the drop (Figure~\ref{Fz_vs_Pe_d}, center). \nd{The fixed drop now pumps fluid towards one of its poles and experiences a force, $F_z$, which can be negative or positive due to the symmetric nature of the bifurcation when $d \to \infty$ (see Fig.~\ref{symm_bif}).} $F_z$ initially increases with increasing $\Pe$ as stronger advection increases the polarity of the surface concentration. \nd{Around $13 < \Pe < 14$, the system exhibits a bi-stability, i.e., depending on the initial conditions, the flow around the swimmer can exhibit either dipolar or quadrupolar symmetry. The latter is characterized by solute-rich regions near both poles and a solute-deficient equator (Figure~\ref{Fz_vs_Pe_d}, right). Correspondingly, $\nabla_sc^b$ changes sign along the drop surface, which leads to a symmetric extensile flow around the drop and a zero net hydrodynamic force $F_z$. Figure~\ref{Fz_vs_Pe_d} shows the bi-stability of the base state beyond $\Pe=13$ \nd{for $d=50$}, with the co-existence of a quadrupolar and a dipolar branch of solutions, for the same values of $\left( d, \Pe \right)$. It can be seen that for $d \to \infty$, the quadrupolar branch corresponds to $F_z \approx 0$.} We stress here that linear stability analysis predicts the emergence of this quadrupolar flow mode, from an initially isotropic state, for $Pe \ge 12$~\cite{Michelin2013}. This is however not incompatible with the present nonlinear simulations that show a definite occurrence of the quadrupolar mode for $\Pe > 13$, as the dipolar state may remain promoted over its quadrupolar counter-part by non-linear interaction of the flow field and solute distribution.

\begin{figure}[h]
    \begin{center}
    \includegraphics[width=5.5cm]{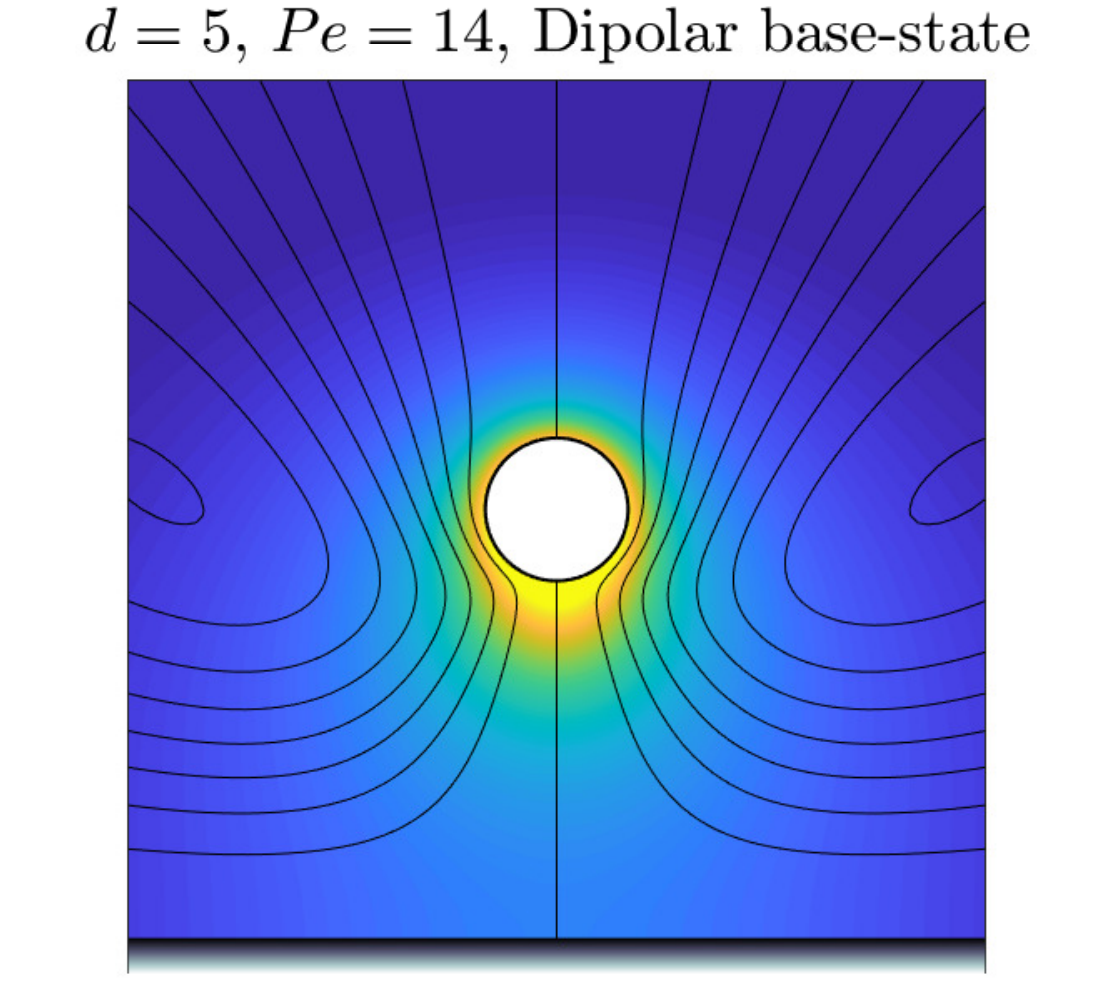}
    \hspace{1cm}
    \includegraphics[width=5.5cm]{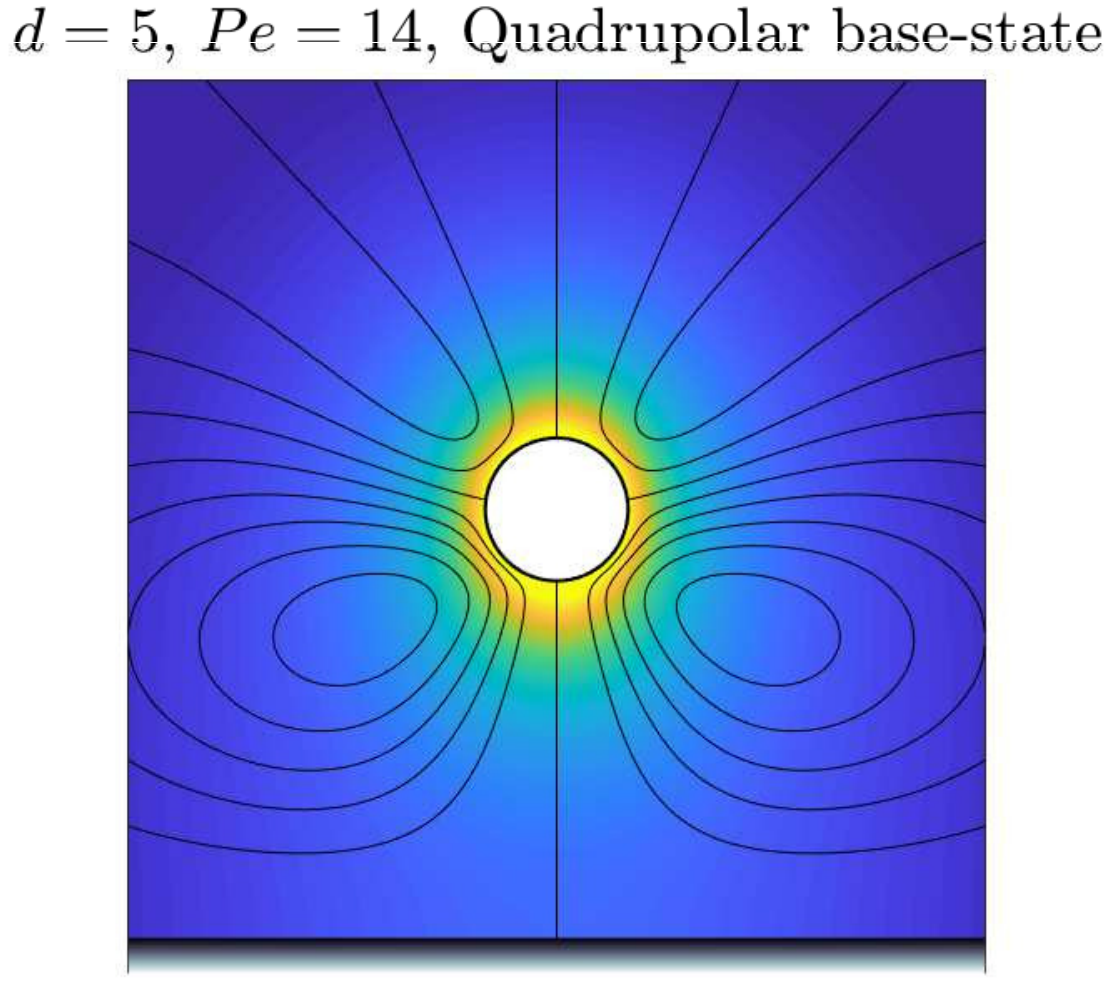}\\
    \caption{Bi-stability of the base state: the concentration distribution and streamlines are shown for the two different branches of solutions co-existing for $d=5$ and $\Pe=14$, showing respectively dipolar (left) and quadrupolar (right) symmetry. The relative stability of these two branches is analysed in Section~\ref{s:res2} and Fig. \ref{adv_term_comp_same_d_quad}, the quadrupolar branch is more unstable to longitudinal swimming modes.}
    \label{two_types_baseState}
    \end{center}
    \end{figure}

\nd{As the drop-wall separation is reduced, the key features of the previous discussion are still observed -- the dipolar flow symmetry for lower $\Pe$, \nd{the dipolar/quadrupolar bi-stability} and the quadrupolar flow symmetry at higher $\Pe$ -- but with important changes due to the increased wall proximity.} The wall prevents downward diffusion of the solute, thus causing higher solute concentration near the bottom pole and polarizing the solute concentration even for $\Pe<8$. The corresponding base states are no longer quiescent and $F_z \ne 0$ (Figure~\ref{Fz_vs_Pe_d}). However, the direction of flow is now conditioned by the wall and we only observe flow toward the wall in our numerical solutions, i.e., the $F_z > 0$ branch of the symmetric bifurcation. This feature suggests the existence of an imperfect bifurcation at $\Pe=8$ in a similar spirit to that identified for the self-propelling mode at $\Pe=4$~\cite{Lippera2020, Saha2021}. \nd{The bi-stability observed for $d \to \infty$ is also present for $d=5$, within similar range of $\Pe$.} A visualisation of the dipolar and the quadrupolar solutions for $d=5$ and $\Pe=14$ is presented in Fig.~\ref{two_types_baseState}. The relative stability properties of these two base states with respect to horizontal swimming modes will be analysed in Section~\ref{s:results}. Finally, beyond $\Pe=14$, $F_z$ is still reduced drastically but it does not drop to zero as in the $d \to \infty$ case. The quadrupolar state from the $d \to \infty$ case is transformed into a `near-quadrupolar' state, with a wall-induced bias toward the bottom pole. This is reflected in the top-down asymmetry of the streamlines, as seen in Figure~\ref{Fz_vs_Pe_d} by comparing the flow profile for $d=5$ and $d=50$ when $\Pe=14$. \nd{More physically, the stagnation point ($\textbf{u} = \textbf{0}$) on the drop surface, which is found at the equator for $d \to \infty$, is shifted upwards when $d$ is finite.}

Understandably, reducing $d$ increases solute accumulation near the bottom pole while solute diffusion away from the top pole is not affected by the wall proximity. \nd{For a fixed P\'eclet number, this results in a monotonic increase in the force experienced by the drop, for both the dipolar and the quadrupolar base states. Now however, the branch corresponding to the quadrupolar mode disappears for lower P\'eclet numbers, resulting in a monotonic increase in the P\'eclet number corresponding to the first occurrence of the quadrupolar base state solutions. For example, while quadrupolar states are seen as early as $\Pe = 14$ for $d=5$, their emergence is delayed to $\Pe = 15$ for $d=3$ and even further to $\Pe = 17$ for $d = 2$.} \nd{The loss of quadrupolar flow structure} is clearly visible in the corresponding flow profiles for $\Pe=14$, as the drop-wall separation is reduced from $d=5$ to $d=1$ (Fig.~\ref{Fz_vs_Pe_d}). \nd{In fact, for $d=1$, we do not obtain a steady state quadrupolar flow structure in the $\Pe$-range considered in our analysis.} As can be seen in the rate of increase of $F_z$ with $\Pe$ in Figure~\ref{Fz_vs_Pe_d}, the solute polarization increases rapidly as $d$ is reduced, resulting in the drop experiencing large phoretic forces. In fact, for $d \ll 1$, $F_z$ diverges as $O(d^{-1})$ and only the flow within the thin gap between the drop and the wall contributes to this force \cite{Yariv2016}.

\section{Linear stability analysis} \label{s:stability}

We now turn to the stability analysis of the base state identified in the previous section, with respect to non-axisymmetric swimming modes (i.e. along the wall). In contrast with the unbounded case, where the base state features no fluid motion for $\Pe \le 8$, the phoretic flow induced by the wall-induced asymmetry of the base solute distribution, which is also responsible for the net vertical hydrodynamic force on the droplet, is expected to  change the details of the advective-coupling-induced instability significantly.  Our objective is to provide physical insight on the effects of the non-quiescent base state on the growth rate of the unstable swimming modes. As part of this analysis, we will also identify the dependence of the critical P\'eclet number for self-propulsion ($\Pe_c$), on the drop's equilibrium distance from the wall.

\subsection{Linearised equations in the bispherical framework} \label{s:stability1}

To analyse the stability of the base state, the different fields $(c,\ub,p)$ are expanded as:
\begin{align}\label{primed_vars}
c\left( \xi, \mu, \phi, t \right) & = c^b\left( \xi, \mu \right) + c'\left( \xi, \mu, \phi \right) e^{\sigma t}, \nonumber \\
\textbf{u}\left( \xi, \mu, \phi, t \right) & = \textbf{u}^b\left( \xi, \mu \right) + \textbf{u}'\left( \xi, \mu, \phi \right) e^{\sigma t}, \nonumber \\
p\left( \xi, \mu, \phi, t \right) & = p^b\left( \xi, \mu \right) + p'\left( \xi, \mu, \phi \right) e^{\sigma t},
\end{align}
where primed variables denote asymptotically small perturbations and $\sigma = \sigma_r + i\sigma_i$ is the complex growth rate of the instability. The translational and rotational velocities are expanded similarly as
\begin{align}\label{primed_vels}
    \textbf{V} & = \textbf{V}'e^{\sigma t}, \qquad\textbf{W}  = \mathbf{\Omega}'e^{\sigma t}.
\end{align}
Without any loss of generality we seek eigenmodes that are symmetric in $y$, so that $\mathbf{V}'=V'_z\eb_z+V'_x\eb_x$ and $\boldsymbol\Omega'=\Omega'_y\eb_y$. Substituting Eqs.~\eqref{primed_vars} into the advection-diffusion equation written in the drop's reference frame, and retaining only the terms that are linear in the perturbation variables, provides:
\begin{equation}\label{gde_stab}
    \sigma c' + \textbf{u}^b \cdot \nabla c' + \left( \textbf{u}' - \textbf{V}' \right) \cdot \nabla c^b = \frac{\nabla^2 c'}{Pe},
\end{equation}
subject to the boundary conditions,
\begin{equation}\label{AdvDiff_stab_bcs}
    {\left. {\textbf{n} \cdot \nabla c'} \right|}_{\mathscr{W}} = 0, {\left. {\textbf{n} \cdot \nabla c'} \right|}_{\mathscr{S}} = 0.
\end{equation}
 It should be noted here that the advection-diffusion equation, Eq.~\eqref{gde_stab}, is now written for a set of axes moving with the drop (i.e. fixed $\xi$ or $\mu$ correspond to moving points). The time derivative, $\partial c/\partial t = \sigma c' e^{\sigma t}$, in Eq.~\eqref{gde_stab} thus denotes the rate of change of $c'$ at a point moving rigidly with the droplet in the (laboratory-fixed) cylindrical coordinate system. In Eq.~\eqref{gde_stab}, $\textbf{u}'$ is however still the fluid velocity in the fixed \textit{laboratory} frame: it satisfies the continuity and Stokes equations, Eqs~ \eqref{continuity} and vanishes far from the droplet and at the wall,
    \begin{equation}\label{BC_wall_nonAxisymm}
        {\left. \textbf{u}' \right|}_{\mathscr{W}} = \textbf{0},
    \end{equation}
    and further satisfies,
    \begin{align}\label{BC_drop_nonAxisymm}
        \textbf{u}' & =  \textbf{V}' + \nabla_s c' + \mathbf{\Omega}' \times \textbf{x}_s, \nonumber \\
        & = V'_z\textbf{e}_z + V'_x\textbf{e}_x + \nabla_s c' + \Omega'_y \textbf{e}_y \times \textbf{x}_s,
    \end{align}
    at the drop's surface. In addition, the horizontal force and torque on the drop must stay zero at all times (self-propulsion):
\begin{equation}\label{force_int}
    F_x = \int_{S} { \textbf{n} \cdot \boldsymbol\sigma' \cdot \textbf{e}_x\;dS } = 0,\qquad 
    T_y = \int_{S} { \textbf{x}_s \times \left( \textbf{n} \cdot \boldsymbol\sigma' \right) \cdot \textbf{e}_y\;dS } = 0,
\end{equation}
where, $\boldsymbol\sigma'$ is the stress tensor in the fluid due to the perturbation flow $\textbf{u}'$. Note that because of its axisymmetric structure, the base state, $\textbf{u}^b$, does not contribute to the force and torque components above. Eqs.~\eqref{gde_stab} to \eqref{force_int} define an eigenvalue problem and we seek for conditions when $\sigma_r > 0$, i.e., exponentially-growing perturbations, leading to self-propulsion of the drop. An important detail here is the condition for the velocity component $V'_z$. This is obtained by substituting $\left( \textbf{u}, p \right)$ from Eqs.~\eqref{primed_vars} into the $z$-component of Eq.~\eqref{force_int_gen},
\begin{equation}\label{fz_int}
    F_z = \int_{S} { \textbf{n} \cdot \boldsymbol\sigma' \cdot \textbf{e}_z\;dS } = 0.
\end{equation}
We will see in Section~\ref{s:stability2} that the solution of the eigenvalue problem involves a modal decomposition of the perturbations in Eqs.~\eqref{primed_vars} as linear sums of appropriate eigenfunctions. Due to the linearity of our analysis, the modes corresponding to the drop's motion along $z$ are uncoupled from those corresponding to the drop's motion along $x$. Therefore, when we are analysing the tendency of the drop to swim along $x$ (i.e., its stability to longitudinal perturbations), the velocity component $V'_z$ drops from the analysis, and so does the need to utilize Eq.~\eqref{fz_int}.

\subsection{Solution methodology: general bi-spherical coordinates} \label{s:stability2}

The solution of this system begins with the expansion of the flow and concentration fields as a series of general, bi-spherical harmonics. The former expansion has one key difference as compared to the axisymmetric flow-field expansions, Eq.~\eqref{u}: instead of expressing $\textbf{u}'$ in the bi-spherical system $\left( \textbf{e}_{\xi}, \textbf{e}_{\mu}, \textbf{e}_{\phi} \right)$, we express it in the cylindrical system $\left( \textbf{e}_{z}, \textbf{e}_{\rho}, \textbf{e}_{\phi} \right)$, but with the different components still expressed in terms of the bi-spherical variables $\left( \xi, \mu, \phi \right)$~\cite{Lee1980}. The expansion of the concentration field on the other hand, is a generalization of the axisymmetric expansion of Eq.~\eqref{c_axisymm} in Section~\ref{s:baseState1b} using bispherical harmonics.

\subsubsection{Flow field} \label{s:stability2a}

The expansions of the various components of $\textbf{u}'$ are:
\begin{align}\label{uprime_z}
    u'_z\left( \xi ,\mu ,\phi  \right)& = \frac{z}{2a}\Gamma^{1/2}\sum\limits_{m=0}^{\infty }{\sum\limits_{n=m}^{\infty} \left[ A^m_n \sinh\left\{ (n+1/2)\lambda\xi \right\} + B^m_n \cosh\left\{ (n+1/2)\lambda\xi \right\} \right] P^m_n \left( \mu \right) \cos \left( m\phi  \right)} \nonumber \\ & + \sum\limits_{m=0}^{\infty }{\sum\limits_{n=m}^{\infty }{C^m_n \sinh \left\{ \left( n+1/2 \right) \lambda \xi \right\}P^m_n \left( \mu \right) \cos \left( m\phi  \right) }},
\end{align}
\begin{align}\label{uprime_rho}
    u'_{\rho }\left( \xi ,\mu ,\phi  \right) & = \frac{\rho}{2a}\Gamma^{1/2}\sum\limits_{m=0}^{\infty }{\sum\limits_{n=m}^{\infty} \left[ A^m_n \sinh\left\{ (n+1/2)\lambda\xi \right\} + B^m_n \cosh\left\{ (n+1/2)\lambda\xi \right\} \right] P^m_n \left( \mu \right) \cos \left( m\phi  \right)} \nonumber \\ & + {{\Gamma }^{1/2}}\sum\limits_{n=1}^{\infty }{\left[ E_{n}^{0}\sinh \left\{ \left( n+1/2 \right)\lambda \xi  \right\}+F_{n}^{0}\cosh \left\{ \left( n+1/2 \right)\lambda \xi  \right\} \right]P_{n}^{1}\left( \mu  \right)} \nonumber \\ & +\frac{1}{2}\sum\limits_{m=1}^{\infty }{\left\{ {{\gamma }_{m}}\left( \xi ,\mu  \right)+{{\chi }_{m}}\left( \xi ,\mu  \right) \right\}\cos \left( m\phi  \right)},
\end{align}
and,
\begin{align}\label{uprime_phi}
    u'_{\phi }\left( \xi ,\mu ,\phi  \right) & = {{\Gamma }^{1/2}}\sum\limits_{n=1}^{\infty }{\left[ G_{n}^{0}\sinh \left\{ \left( n+1/2 \right)\lambda \xi  \right\}+H_{n}^{0}\cosh \left\{ \left( n+1/2 \right)\lambda \xi  \right\} \right]P_{n}^{1}\left( \mu  \right)} \nonumber \\ & +\frac{1}{2}\sum\limits_{m=1}^{\infty }{\left\{ {{\gamma }_{m}}\left( \xi ,\mu  \right)-{{\chi }_{m}}\left( \xi ,\mu  \right) \right\}\sin \left( m\phi  \right)},
\end{align}
where, the functions $\gamma_m \left( \xi, \mu \right)$ and $\chi_m \left( \xi, \mu \right)$ in Eqs.~\eqref{uprime_rho} and \eqref{uprime_phi} are given by:
\begin{align}\label{gamma_m_chi_m}
{{\gamma }_{m}}\left( \xi ,\mu  \right) & = {{\Gamma }^{1/2}}\sum\limits_{n=m+1}^{\infty }{\left[ E_{n}^{m}\sinh \left\{ \left( n+1/2 \right)\lambda \xi  \right\}+F_{n}^{m}\cosh \left\{ \left( n+1/2 \right)\lambda \xi  \right\} \right]P_{n}^{m+1}\left( \mu  \right)}, \nonumber \\ 
{{\chi }_{m}}\left( \xi ,\mu  \right)& = {{\Gamma }^{1/2}}\sum\limits_{n=m-1}^{\infty }{\left[ G_{n}^{m}\sinh \left\{ \left( n+1/2 \right)\lambda \xi  \right\}+H_{n}^{m}\cosh \left\{ \left( n+1/2 \right)\lambda \xi  \right\} \right]P_{n}^{m-1}\left( \mu  \right)}.  
\end{align}

\nd{$P^m_n \left( \mu \right)$ refer in Eqns.~\eqref{uprime_z}--\eqref{gamma_m_chi_m} to the associated Legendre polynomials of degree $n$ and order $m$.} The expansions above already satisfy the wall impermeability automatically, ${u_z'(\xi=0)=0}$. The coefficients $\left[ A^m_n...H^m_n \right]$ are obtained by applying the continuity equation, the tangential velocity boundary conditions on the wall, and the normal and tangential velocity boundary conditions on the drop surface. The coefficients corresponding to the $m=1$ mode are the first non-axisymmetric ($\phi$-dependent) contributions to the flow field. Interestingly, this is the only mode responsible for both translation along the $x$-axis with velocity $V'_x$, and rotation about the $y$-axis with velocity $\Omega'_y$. This can be established by evaluating the integrals in Eqs.~\eqref{force_int} using Eqs.~\eqref{uprime_z}--\eqref{gamma_m_chi_m}, \nd{and noting that only} the coefficients of the $m=1$ mode contribute to $F_x$ and $T_y$ (see also Eqs.~\eqref{force_disc} and \eqref{torque_disc}; Refs.~\cite{Lee1980, Mozaffari2016}).

\subsubsection{Concentration field} \label{s:stability2b}

The concentration field, $c'$, is written as:
\begin{equation}\label{c_nonAxisymm}
    c'\left( \xi, \mu, \phi  \right)= \Gamma^{1/2} \sum\limits_{m=0}^{\infty } { \sum\limits_{n=0}^{\infty }{c^m_n \left( \xi  \right) P^m_n\left( \mu  \right) \cos \left( m\phi  \right)} },
\end{equation}
where $c^m_n \left( \xi \right)$ are unknowns eigenfunctions  to be determined. Once again, the coefficients corresponding to the $m=1$ mode are the first non-axisymmetric contributions to the concentration field, and are the only ones corresponding the correct polar symmetry of a self-propelling mode along $\eb_x$. \nd{The axisymmetry ($\phi$-independence) of the base state means that the nonlinear (convective) terms in Eq.~\eqref{gde_stab} do not introduce any coupling between modes of different $m$ order: a projection onto $\cos (\phi)$ of Eq.~\eqref{gde_stab} results in all the $m > 1$ polar modes (i.e., $\cos (m\phi)$ terms for $m > 1$) dropping from the analysis. As a result, the $m=1$ concentration modes are not coupled to the $m > 1$ concentration modes. In addition, the linearity of the hydrodynamics problem means that the $m=1$ velocity modes are directly and only related to the $m=1$ concentration modes and are therefore decoupled from  the $m > 1$ velocity modes.} Therefore, we only need to solve for the $m=1$ modes of the expansions in Eqs.~\eqref{uprime_z} to \eqref{uprime_phi} and Eq.~\eqref{c_nonAxisymm} to completely characterize the onset of advection-induced spontaneous swimming.

\subsubsection{Projected equations} \label{s:stability2c}

The following analysis is thus purposely restricted to $m=1$. \nd{Similar to the analysis of} the axisymmetric case ($m=0$), the flow field components, or equivalently $\left\{A^1_1,...,H^1_N\right\}$, can be directly and linearly expressed in terms of the surface concentration components $c_n^1(\xi=1)$. To do so, the boundary condition Eq.~\eqref{BC_drop_nonAxisymm} is projected onto the appropriate Legendre modes. The system of linear equations is then completed by projection of the continuity condition for the flow field, as well as the wall boundary condition, Eq.~\eqref{BC_wall_nonAxisymm} and force-free and torque-free conditions, Eq.~\eqref{force_int}. This is a classical analysis in general bi-spherical coordinates (see \cite{Lee1980, Mozaffari2016}) and is described in the Appendix.

Additionally, substitution of the various perturbation fields, Eqs.~\eqref{uprime_z}--\eqref{c_nonAxisymm}, and base state solutions (Section~\ref{s:baseState}) into Eq.~\eqref{gde_stab}, and projection of the resulting transport equation (after division by $\Gamma$) provide successive linear, coupled differential equations for the different concentration modes $c_n^1(\xi)$:
\begin{equation}\label{AdvDiff_stab_disc}
    \sigma \mathbf{H}^2 \cdot \mathbf{C}' + \mathbf{B}^4 \cdot \mathbf{C}' + \mathbf{B}^5 \cdot \frac{d \mathbf{C}'}{d \xi } + \mathbf{B}^6 \cdot \mathcal{U}_H + \mathbf{B}^7 V'_x = \frac{1}{Pe}\left\{ {{\mathbf{A}}^{3}}\cdot \mathbf{C}' + {{\mathbf{A}}^{4}}\cdot \frac{d^2 \mathbf{C}'}{d \xi ^2} \right\},
\end{equation}
with  $\textbf{C}' \equiv \left[ c^1_1(\xi),\;c^1_2(\xi),\;...,\;c^1_N(\xi) \right]$, $\mathcal{U}_H \equiv \left[ A^1_1,...,H^1_N \right]$ and where the tensors $\textbf{H}^2$, $\textbf{A}^i$ and $\textbf{B}^i$ depend only on $\xi$ (see Appendix). The second order tensors $\textbf{B}^4$ and $\textbf{B}^5$ result from the projection of $\mathbf{u}^b \cdot \nabla c'$, and contain information about the base state flow. The $\left( \textbf{B}^6 \cdot \mathcal{U}_H \right)$ term abbreviates the projection of $\mathbf{u}' \cdot \nabla c^b$, and contains information about the base state solute distribution. The expanded version of this term is detailed in Eq.~\eqref{disc_uprime_grad_cb} in the Appendix. Finally, the $\left( \mathbf{B}^7 V'_x \right)$ term results from the projection of $-\mathbf{V}' \cdot \nabla c^b$, and is related to the distortion of $c^b$ due to the drop's motion along the wall. Associated boundary conditions for $c^1_n(\xi)$ come from the projections of Eqs.~\eqref{AdvDiff_stab_bcs} onto $P^1_k(\mu)\cos(\phi)$, which yield:
\begin{equation}\label{AdvDiff_NonaxiSymm_disc_BCs1}
    \left. \left( \frac{d c^1_k}{d \xi} - \frac{k+2}{2k+3} \frac{d c^1_{k+1}}{d \xi} -\frac{k-1}{2k-1} \frac{d c^1_{k-1}}{d \xi} \right) \right|_{\xi=0} = 0,
\end{equation}
and,
\begin{equation}\label{AdvDiff_NonaxiSymm_disc_BCs2}
    \left. \left( \frac{\lambda \sinh(\lambda)}{2}c^1_k + \cosh(\lambda) \frac{d c^1_k}{d \xi} - \frac{k+2}{2k+3} \frac{d c^1_{k+1}}{d \xi} - \frac{k-1}{2k-1} \frac{d c^1_{k-1}}{d \xi} \right) \right|_{\xi=1} = 0,
\end{equation}
with $1 \le k \le N$ for both Eqs.~\eqref{AdvDiff_NonaxiSymm_disc_BCs1}--\eqref{AdvDiff_NonaxiSymm_disc_BCs2}.

This problem is solved numerically by discretising the different functions $c_n^1(\xi)$ over $M$ equispaced points for $0\leq \xi\leq 1$, so that the solute transport problem as $N\times M$ unknowns. The solution methodology then follows a similar approach to that of Section-\ref{s:baseState1}, with the fundamental difference that the problem is now linear in the perturbation quantities which are the only unknowns of the problem:
\renewcommand{\labelenumi}{\roman{enumi}}
\begin{enumerate}
    \item for given $\mathbf{C}'$, solve for the hydrodynamics problem by expliciting a formal dependence of the fluid velocity coefficients $\mathcal{U}_H$ on the surface concentration modes,
    \item using the previous step, express the solute transport problem in terms of $\textbf{C}'$ alone, Eqs.~\eqref{gde_stab}--\eqref{AdvDiff_stab_bcs}.
\end{enumerate}

After appropriate discretization of the $\xi$-derivatives of $\mathbf{C}'$ (in the present work, we use second-order accurate, centered finite differences), the projected hydrodynamic and solute transport problems can be assembled into a linear eigenvalue problem:
\begin{equation}\label{EVP_matrix_form}
    \begin{bmatrix}
    \textbf{A}_{11} & \textbf{B}^6 & \textbf{A}_{13}\\
    \textbf{A}_{21} & \textbf{A}_{22} & \textbf{A}_{23}\\
    \textbf{0} & \textbf{A}_{32} & \textbf{0}\\
    \end{bmatrix} \cdot \begin{bmatrix}
    \textbf{C}'\\
    \mathcal{U}_H\\
    \mathcal{V}\\
    \end{bmatrix} = \sigma \begin{bmatrix}
    \textbf{H}^2 & \textbf{0} & \textbf{0}\\
    \textbf{0} & \textbf{0} & \textbf{0}\\
    \textbf{0} & \textbf{0} & \textbf{0}\\
    \end{bmatrix} \cdot \begin{bmatrix}
    \textbf{C}'\\
    \mathcal{U}_H\\
    \mathcal{V}\\
    \end{bmatrix},
\end{equation}
with $\mathcal{V} \equiv \left[ V'_x,\;\Omega'_y \right]^T$. The tensors $\textbf{A}_{ij}$ are obtained through the projection of the transport problem evaluated at discrete points (first row), and the projection of the hydrodynamic problem (last two rows) -- see Eqs.~\eqref{AdvDiff_NonaxiSymm_disc_BCs1}, \eqref{AdvDiff_NonaxiSymm_disc_BCs2} and \eqref{disc_sigma_cprime}--\eqref{disc_U_grad_cb}.

\subsection{Validation} \label{s:res1}

\begin{figure}[ht]
\begin{center}
      \includegraphics[width=8cm]{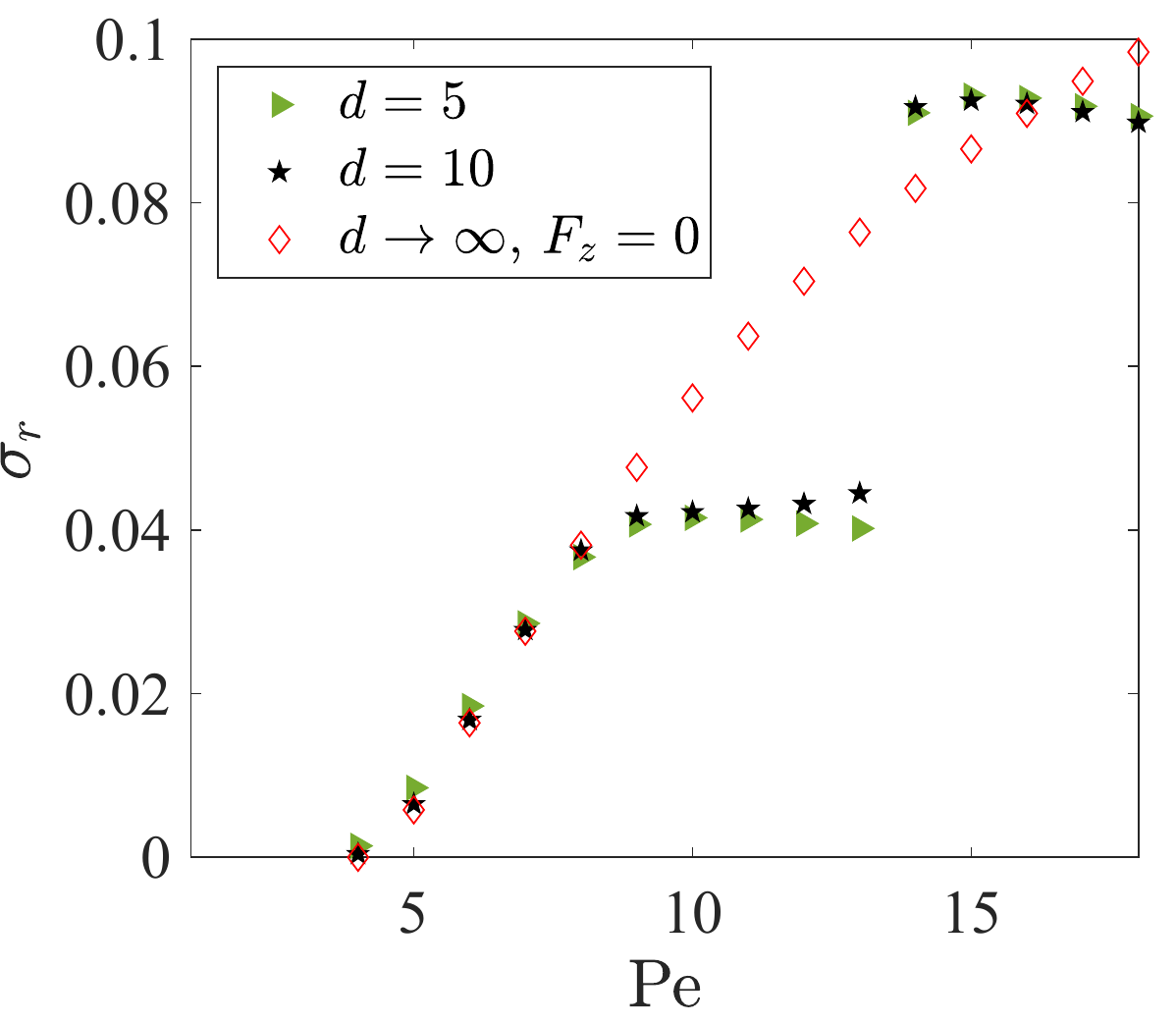}
  \caption{Evolution of the growth rate of the unstable swimming mode for levitating droplets (forced $F_z\ne 0$, $F_x=F_y=0$) for drop-wall separations $d=5$ and $d=10$. The results for a force-free particle in an unbounded fluid (i.e., $d \to \infty$, $\mathbf{F}=\textbf{0}$) are shown for comparison.}
  \label{sigma_vs_Pe_large_d}
\end{center}
\end{figure}

The base state solution for $d \to \infty$ \textit{and} $\Pe < 8$ is essentially the isotropic solution for a stationary force-free phoretic particle in a quiescent fluid (Section~\ref{s:baseState2}), and its stability was investigated by Ref.~\cite{Michelin2013}. Our results for the growth rate, $\sigma_r$, of the unstable modes in that limit are in very good agreement with those of Ref.~\cite{Michelin2013}, up until $\Pe=8$ \nd{(see Fig. \ref{sigma_vs_Pe_large_d}, for $d = 10$)}. \nd{We also recover the value of the critical P\'eclet number for self-propulsion in the unbounded case, $\Pe_c(d = 10) \approx 4$, thus validating the present numerical implementation.}

\section{Wall effects on the drop stability and self-propulsion} \label{s:results}
\subsection{Stability of the nearly-isotropic base state and self-propulsion}
\nd{The first effects of the wall can be seen in Fig. \ref{sigma_vs_Pe_large_d}, most clearly for $d=5,\;\Pe = 5$, in the form of a slight increase in $\sigma_r$. This means that the wall enhances the tendency of the active drop to swim horizontally. It also means that the critical P\'eclet number for self-propulsion parallel to the wall (i.e., the $\Pe$ corresponding to $\sigma_r \approx 0$) is \textit{slightly lowered} when $d=5$. Indeed $\Pe_c \approx 3.56$ for $d=5$, as compared to its value $\Pe_c^{\infty} = 4$ when $d \to \infty$. However, we note that the magnitude of the growth rates does not vary too strongly for $\Pe < 8$, even for drop-to-wall-distances as small as $d=5$. Thus the drop's stability to longitudinal swimming modes is not very significantly impacted when the anisotropy of the base state is induced solely by moderate proximity to a rigid wall ($d\gtrsim 5$).} The fundamental physical mechanisms at the heart of the onset of propulsion of the droplet therefore remain that of the unbounded case and are analysed below.

 The evolution equation for the concentration perturbation, $c'$, is given by:
\begin{equation}\label{dcdt}
    \frac{\partial c'}{\partial t} = - \textbf{u}' \cdot \nabla c^b - \textbf{u}^b \cdot \nabla c' + \textbf{V}' \cdot \nabla c^b + \frac{1}{Pe} \nabla^2 c'.
\end{equation}
To overcome the homogenising effect of diffusion and maintain an asymmetric solute distribution on the drop-surface, $\nabla_s c' \ne 0$, solute transport by the flow must play a critical role and are therefore the object of our focus to understand the origin of the instability.

\begin{figure}[t]
    \begin{center}
    \includegraphics[width=\linewidth]{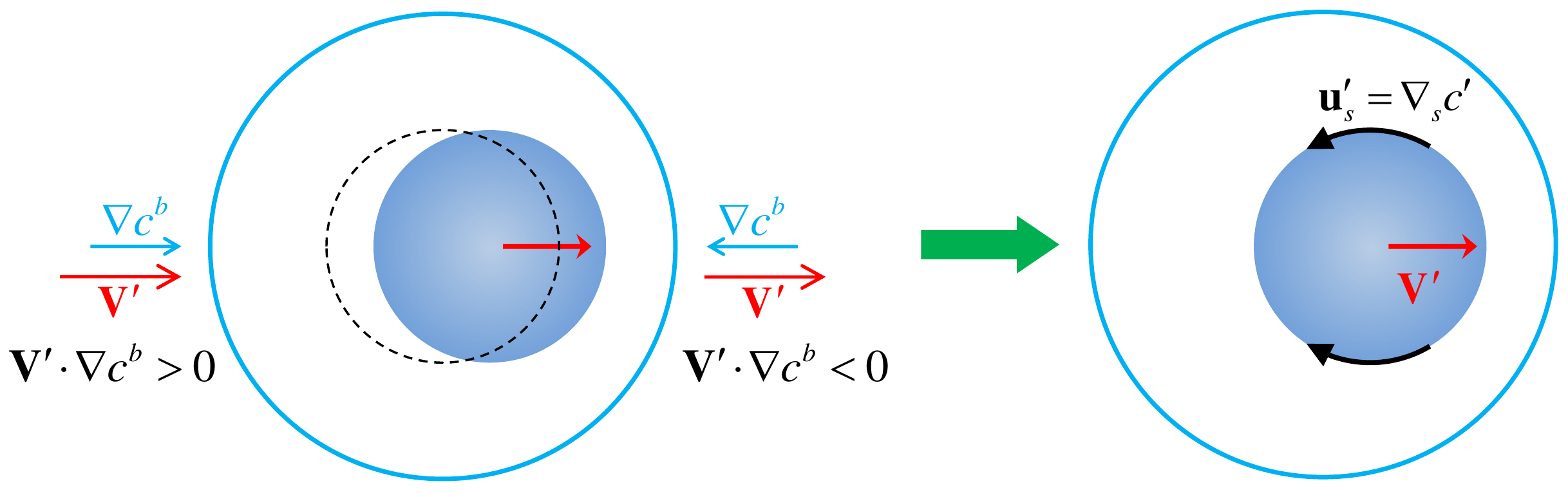}
    \caption{Instability mechanism for $\Pe<8$ in the far-field interaction limit, $d \gg 1$. Left: An initially isotropic solute distribution becomes asymmetric due to a small translation of the drop from its initial position (dashed). A representative isoline of solute concentration is materialised by the blue circle. The resulting concentration perturbations, $c'$, depend on the local sign of $\textbf{V}' \cdot \nabla c^b$ (see Eq.~\eqref{dcdt}). Right: The fore-aft asymmetry in solute concentration leads to the surface slip $\textbf{u}'_s = \nabla_s c'$, which is maintained if advection is strong, causing the drop to swim spontaneously (see Eq.~\eqref{U_reciproc_unb}).}
    \label{adv_term_comp_d5_Pe6}
    \end{center}
    \end{figure}

 The first two terms on the right-hand-side can be combined into $- 2\textbf{u}^b \cdot \nabla c'$ at the drop surface using the boundary conditions \eqref{Stokes_bcs_drop} and \eqref{BC_drop_nonAxisymm}. Furthermore, the fluid is at rest in the base state when $d \to \infty$ and $\Pe < 8$, so that only the interaction of the imposed drop velocity with the base state solute distribution (i.e., the $\textbf{V}' \cdot \nabla c^b$) would be able to amplify asymmetries in solute concentration, and thus to lead to self-propulsion. As the drop is displaced due to an imposed fluctuation velocity, $\textbf{V}' = V'_x \textbf{e}_x$, the resulting perturbed solute field is asymmetric due to the solute accumulation behind the particle, and the associated slip flow is oriented toward the back of the particle, thus sustaining/reinforcing the drop's motion (Fig. \ref{adv_term_comp_d5_Pe6}).

\subsection{Stability of asymmetric base states}

We re-emphasize that for $d \to \infty$ and $\Pe=8$, a bifurcation  occurs in the base state: the stationary drop is now `pumping' fluid axi-symmetrically and experiences a net hydrodynamic force $\textbf{e}_z$ that must be balanced by an external force (e.g., gravity). Thus, for $d \to \infty$ and $\Pe > 8$, we are investigating the stability$-$to longitudinal perturbations$-$of a stationary drop in a regime that is quite different from the study of \cite{Michelin2013}, and results are not able to match that limit anymore. Indeed, for a given $\Pe$, the growth rate is now reduced in comparison to the stability results of the isotropic solution~\cite{Michelin2013}. Some insights on the relative stability of the asymmetric and isotropic base states can be obtained by expressing the velocity of the drop, $\textbf{V}'$, directly in terms of the surface gradient of the concentration, $\nabla_s c'$, using the reciprocal theorem:
\begin{equation}\label{U_reciproc}
    \int_{S} { \textbf{n} \cdot \boldsymbol{\hat\sigma} \cdot \nabla_s c'\;dS } = \textbf{e}_x \cdot \textbf{V}',
\end{equation}
where $\boldsymbol{\hat\sigma}$ is the stress tensor of the auxiliary Stokes flow problem corresponding to the motion of a rigid sphere under the effect of a unit force along the $x$-direction (hence the $\textbf{e}_x$ in the right side of Eq.~\eqref{U_reciproc}) and no net torque. When hydrodynamic interaction with the wall is negligible (e.g. in an unbounded fluid), $\boldsymbol{\hat\sigma}\cdot\mathbf{n}$ is uniform at the particle surface, so that Eq.~\eqref{U_reciproc} simplifies to~\cite{Michelin2013},
\begin{equation}\label{U_reciproc_unb}
    \textbf{V}' = -\frac{1}{4\pi} \int_{S} { \nabla_s c'\;dS } = -\frac{1}{2\pi} \int_{S} { c'\textbf{n}\;dS }.
\end{equation}
In the absence of any hydrodynamic interaction swimming velocity is therefore proportional to the first moment (polarity) of the surface concentration perturbation. This quantity is expected to be relatively larger when the base state is isotropic than for the anisotropic and forced base state. Indeed, for the latter, the non-zero base-flow, $\textbf{u}^b$, `pushes' the solute toward the bottom pole of the drop, causing significant deficit in solute concentration in other regions and driving down the first moment of $c$. A smaller swimming speed resulting from the mechanism described above is indicative of lower instability growth rates, which then explains the reduced growth rates of the non-quiescent (also anisotropic, `forced') base state in this study, w.r.t. the quiescent (also isotropic, force-free) base state considered in Ref.~\cite{Michelin2013}.\\

\begin{figure}[t]
    \begin{center}
    \includegraphics[width=6cm]{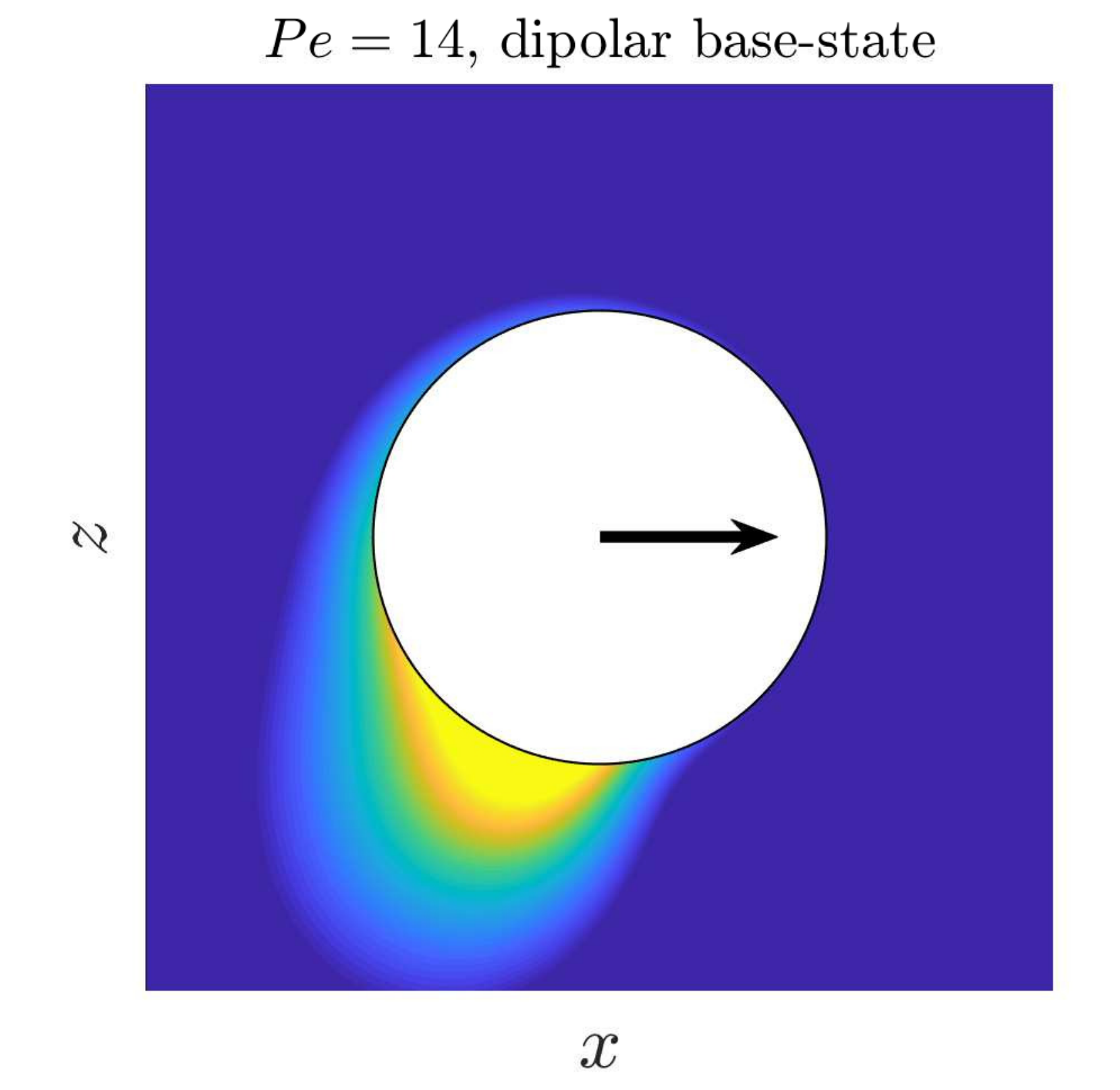}
    \includegraphics[width=6cm]{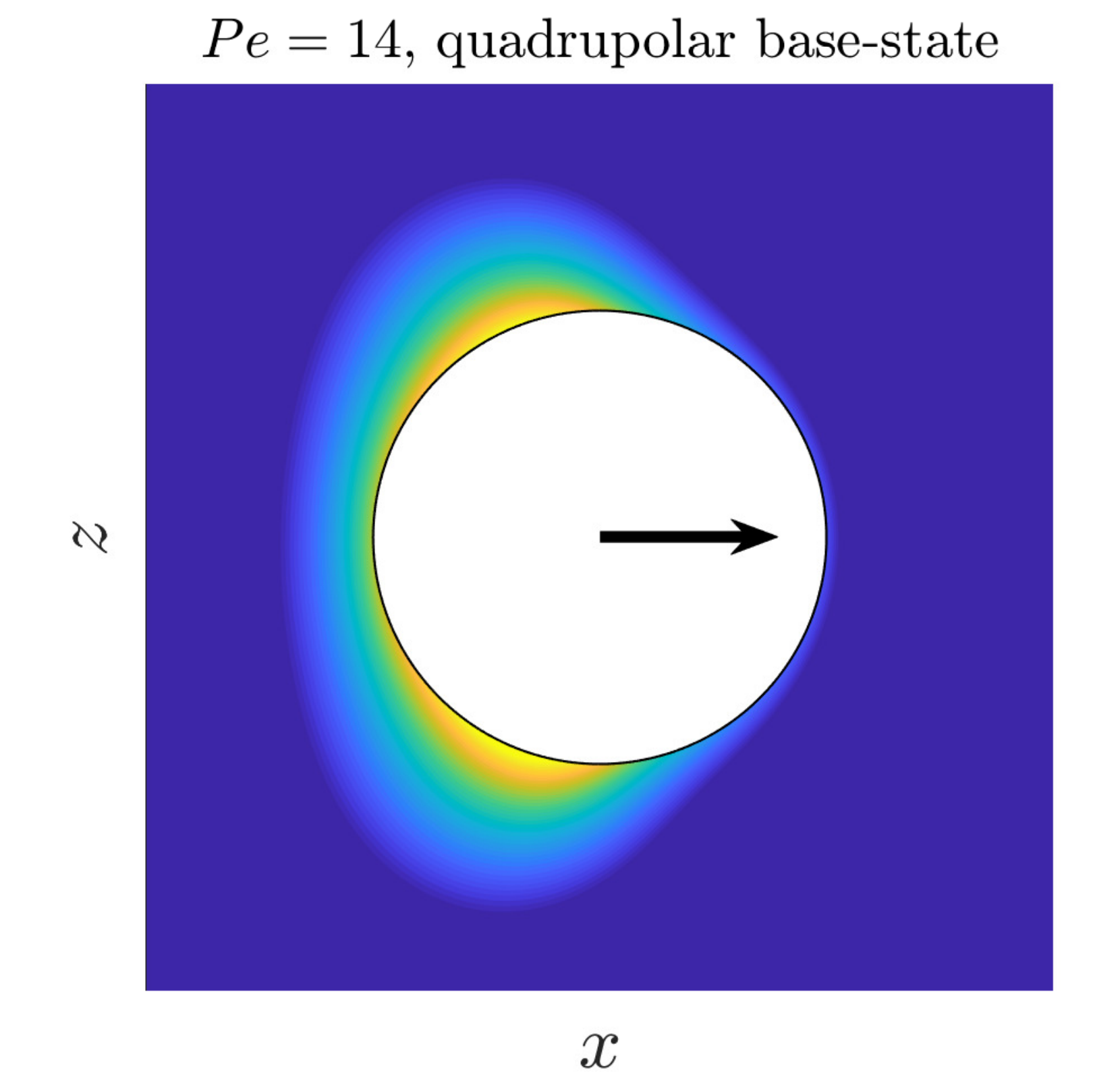}
    \caption{Higher instability of the quadrupolar base state than the dipolar base state, for a horizontally swimming drop (denoted by the black arrow). The contours are qualitative depictions of the solute distribution around the drop in the self-propelling state, with lighter (resp. darker) colors denoting higher (resp. lower) concentrations.}
    \label{adv_term_comp_same_d_quad}
    \end{center}
    \end{figure}

A second bifurcation occurs at $\Pe \approx 13$, wherein the base state flow becomes quadrupolar with solute-rich regions near both the poles of the drop. In terms of the structure of the flow (and the solute concentration), this is very different from the dipolar symmetry observed for $\Pe < 13$ (see Fig.~\ref{Fz_vs_Pe_d}). A marked difference in the base state thus results in a discontinuity in the growth rate w.r.t. $\Pe$ with a relatively large increase in $\sigma_r$ from $\Pe=13$ to $\Pe=14$ as seen in Fig. \ref{sigma_vs_Pe_large_d}, indicating that quadrupolar base states are more prone to the instability than dipolar base states. This can again be explained by examining the nature of concentration distribution around the drop and relating it to the swimming speed via Eq.~\eqref{U_reciproc_unb}. The qualitative solute distributions around these base states, when they are subjected to a longitudinal perturbation, are shown in Fig.~\ref{adv_term_comp_same_d_quad}; where the drop is swimming along $\eb_x$. \nd{For the dipolar base state, the drop is experiencing a constant \textit{external} force normal to the direction of motion. This balances the concentration polarity along $\eb_z$ and ensures that the drop swims along $\eb_x$ (Fig.~\ref{adv_term_comp_same_d_quad}, left). For the quadrupolar base state, there is no external force normal to the direction of motion (for $d \to \infty$), since the $z$ polarity is identically zero (Fig.~\ref{adv_term_comp_same_d_quad}, right).} The dipolar state has strong solute accumulation only near the bottom pole, while the quadrupolar state has enhanced solute concentration near \textit{both} poles. The markedly higher \textit{surface} concentration in the latter case results in larger polarity along $\eb_x$ (see Eq.~\eqref{U_reciproc_unb}), leading to enhanced propulsion/faster destabilization.

\subsection{Wall effects on drop stability} \label{s:res2}

    \begin{figure}[t]
    \begin{center}
    \subfloat[]{\label{sigma_vs_Pe_d_a}\includegraphics[width=7.5cm]{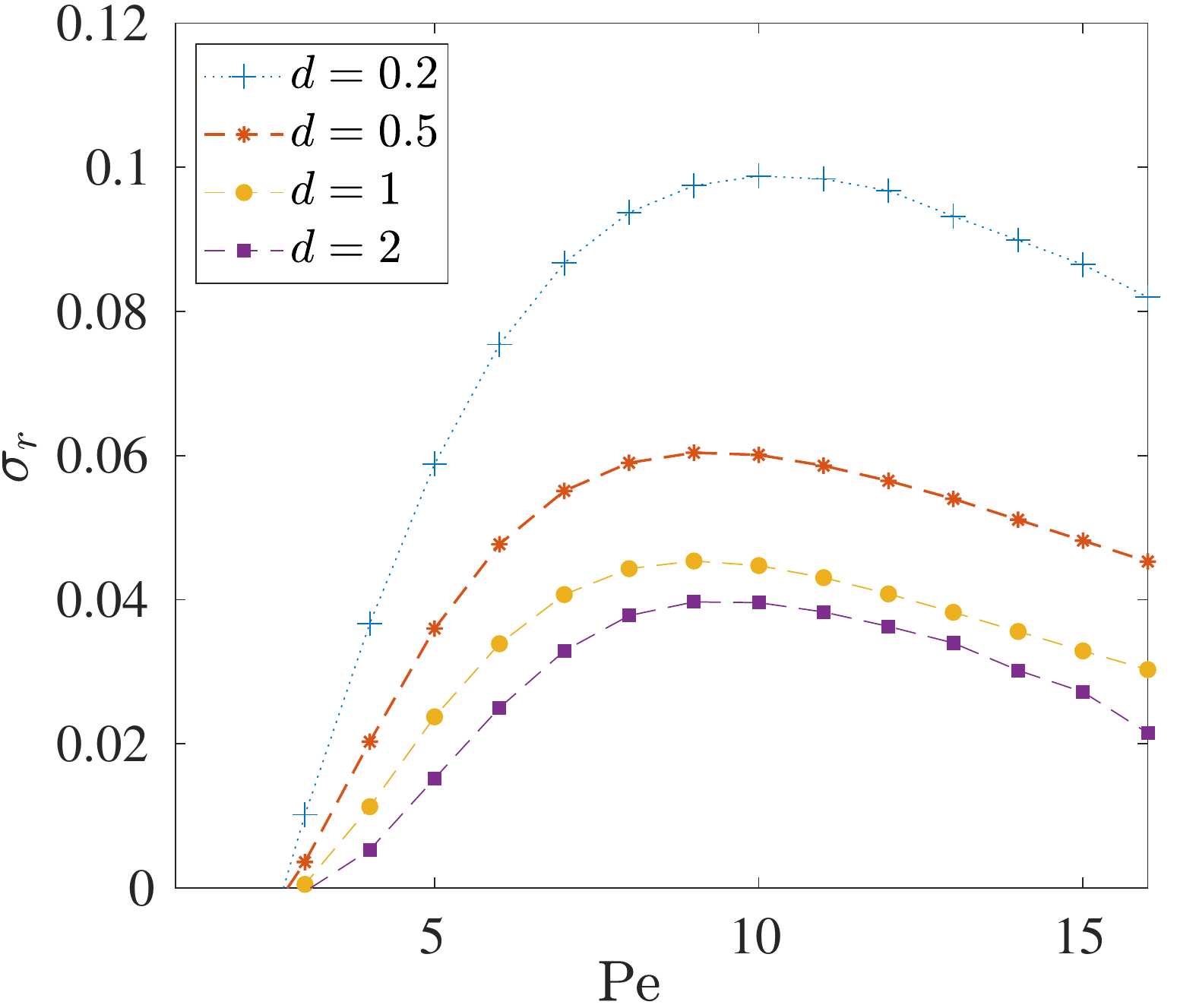}}
    \subfloat[]{\label{sigma_vs_Pe_d_b}\includegraphics[width=7.5cm]{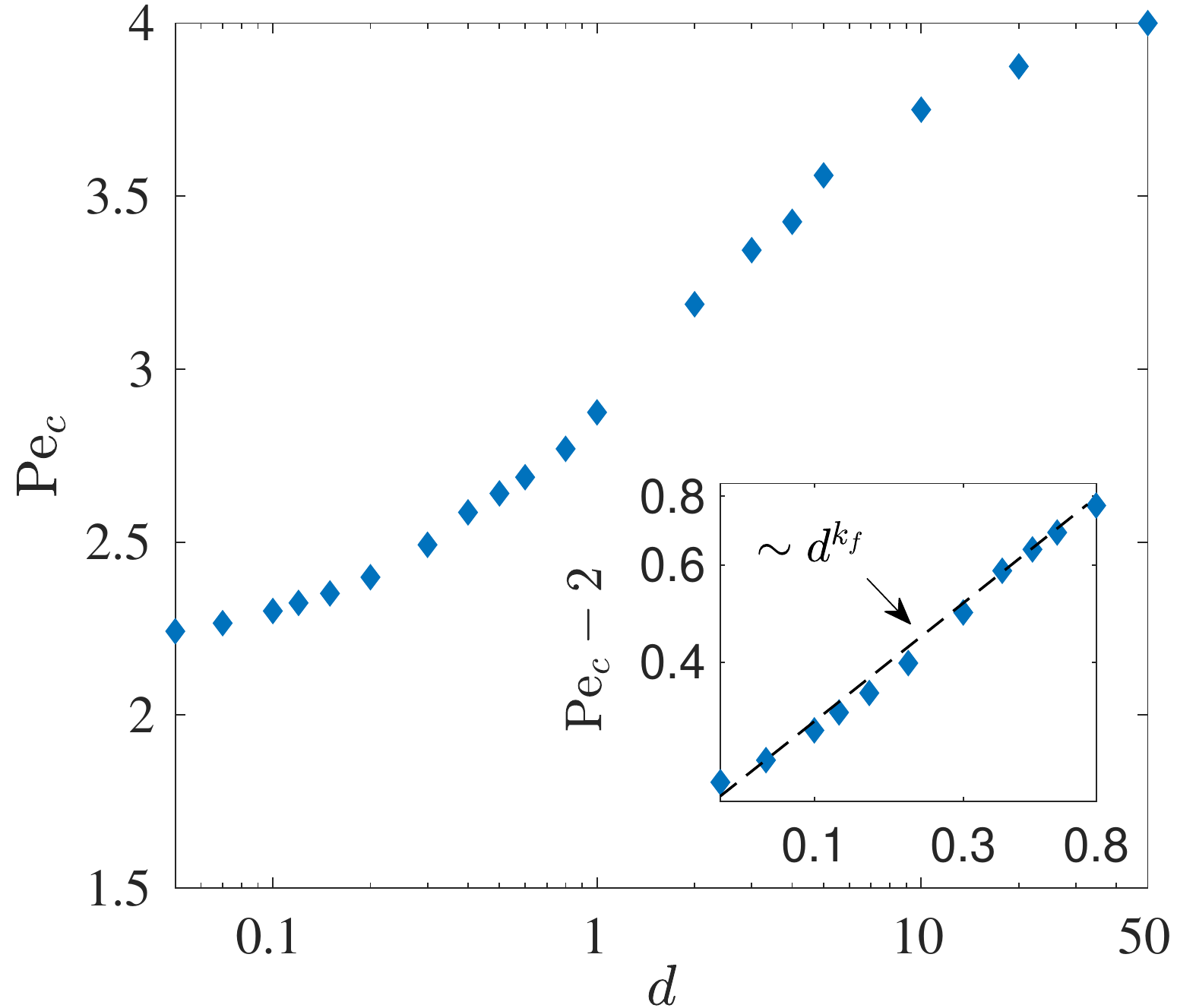}}

    \caption{(a) Evolution with $\Pe$ of the growth rate, $\sigma_r$, of the unstable horizontal swimming mode for small-to-moderate wall distance, $d$. (b) Evolution with $d$ of the critical P\'eclet number for the onset of self-propulsion, $\Pe_c$. The log-log inset shows the asymptotic dependence of $\Pe_c$ in the lubrication limit $d\rightarrow 0$, where the dashed line is a power-law fit, \nd{$\sim d^{k_f}$, $k_f \approx 0.46$}.}
    \end{center}
    \end{figure}

As the drop-wall separation is reduced, the base state corresponding to quadratic flow is suppressed and the growth rate curves become continuous (see Figs.~\ref{sigma_vs_Pe_d_a} and \ref{sigma_vs_Pe_large_d}). For fixed $\Pe$, the growth rate $\sigma_r$ is now increasing monotonically as $d$ is reduced. \nd{A direct consequence is the monotonic reduction, w.r.t. $d$, of the critical P\'eclet number for wall-parallel propulsion, as shown in Fig.~\ref{sigma_vs_Pe_d_b}.} To explain \nd{the wall effects for small values of $d$}, we turn once again  to the relationship between surface concentration and droplet velocity obtained from the reciprocal theorem. Equation~\eqref{U_reciproc} remains valid but the auxiliary stress distribution at the surface (which plays the role of an influence function for the phoretic slip in Eq.~\eqref{U_reciproc}) is not uniform anymore, indicating that the same hydrodynamic slip magnitude will not have the same effect on propulsion depending on its exact location around the droplet. 

\begin{figure}[t]
    \begin{center}
    \includegraphics[width=4.2cm]{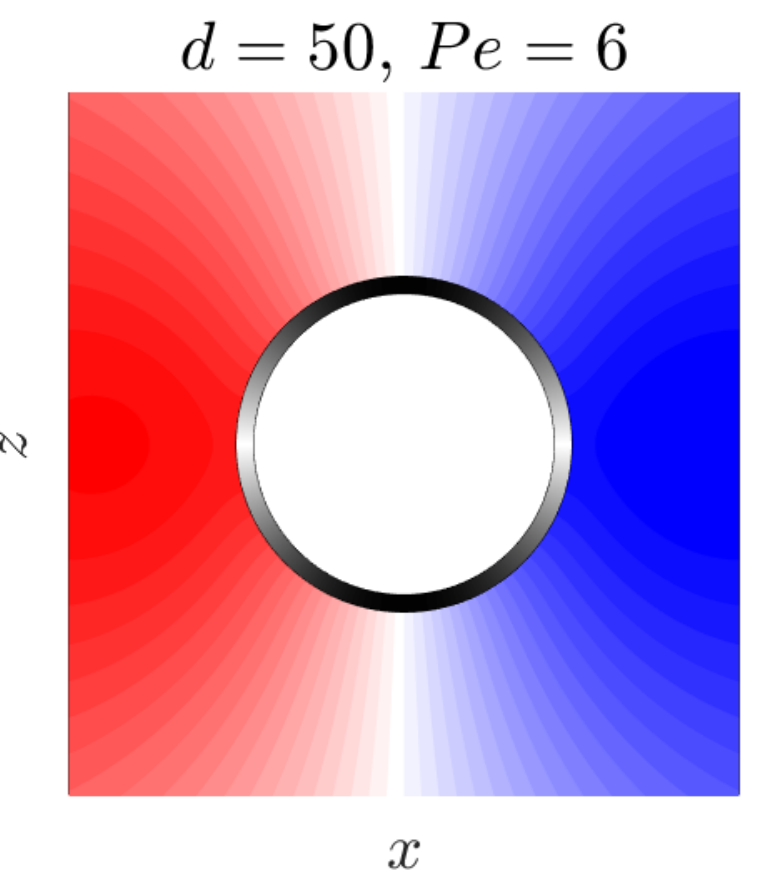}
    \vspace{0.4cm}
    \includegraphics[width=\linewidth]{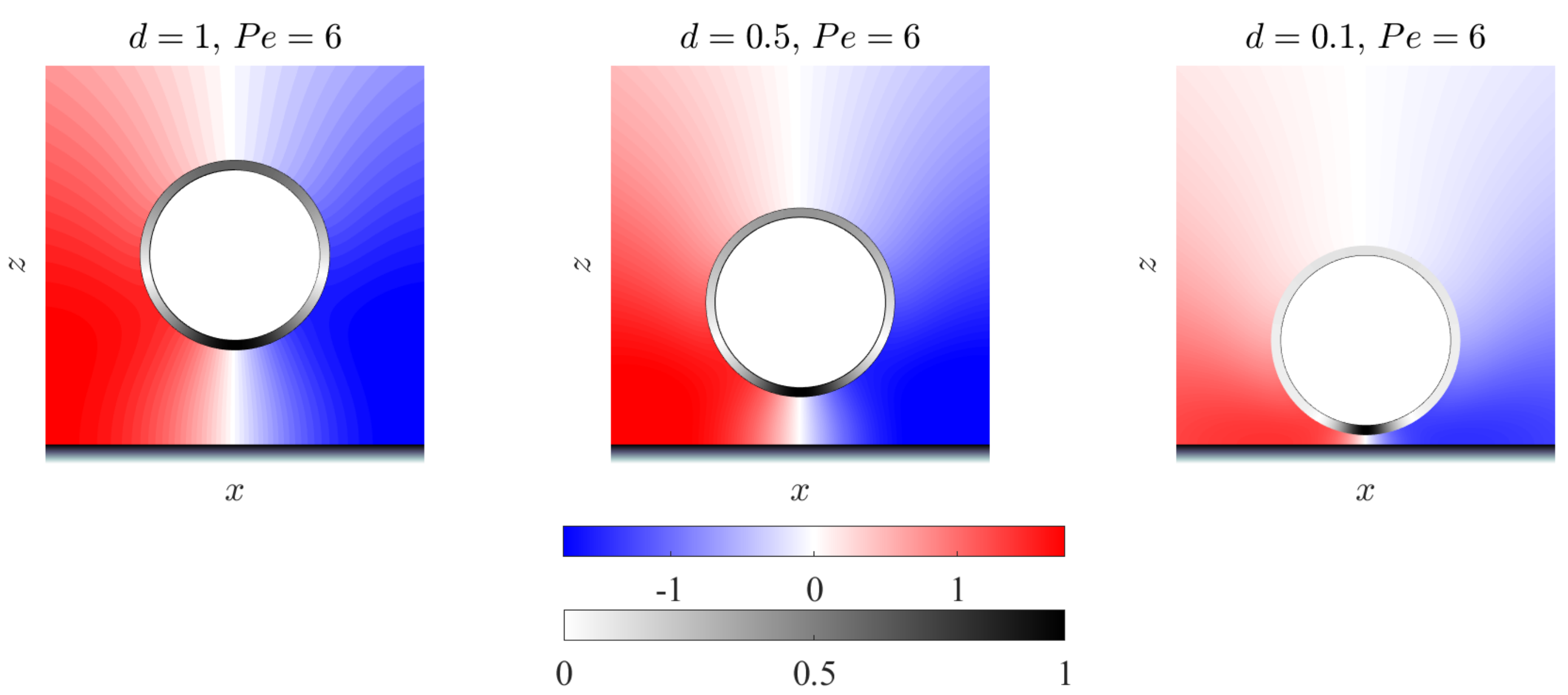}\\
    \caption{Evolution with $d$ of the perturbation concentration $c'$ around the drop normalised by the corresponding swimming speed (color) and normalised auxiliary surface traction $|\mathbf{n}\cdot\boldsymbol{\hat\sigma}\cdot\mathbf{e}_\theta|$ (black and white). A $x-z$ cross section of the distributions is shown. The drop is translating toward the $+ \textbf{e}_x$ direction leading to a solute deficiency in its direction of motion. Here, $\textbf{e}_{\theta}$ is the tangent unit vector on the drop's surface with $\theta$ measured counter-clockwise from $\textbf{e}_{x}$. The top row shows these same quantities in the limit $d \to \infty$, and a clear symmetry w.r.t. the direction of motion, $\textbf{e}_x$, is observed.}
    \label{tracn_comp_same_Pe}
    \end{center}
    \end{figure}

As $d$ is reduced, the concentration in the confined region between the drop and the wall is enhanced due to reduced diffusion. In addition, the base flow, $\textbf{u}^b$, drives the concentration disturbances toward the wall, thus generating strong solute concentration gradients, $\nabla_s c'$, in the gap between the wall and the drop. Specifically, $\left| \nabla_s c' \right|$ is greatest near the bottom pole of the drop, where the distance to the wall is smallest. This is also the region where the towed sphere of the auxiliary Stokes problem experiences the most shear, and the slip influence, $\textbf{n} \cdot \boldsymbol{\hat\sigma}$, is therefore strongest (Figure~\ref{tracn_comp_same_Pe}). A combination of these two effects explains the large instability growth-rates for small values of $d$, in Fig.~\ref{sigma_vs_Pe_d_a}. In short, the wall-induced base flow tends to concentrate efficiently the fluid slip where it is expected to generate stronger propulsion, or alternatively, enhance the tendency of the drop to destabilize and swim.

Figure~\ref{tracn_comp_same_Pe} suggests that the transport in the gap plays a dominant role on the drop-dynamics as $d$ is reduced. This is also confirmed by our results where we saw the rapid saturation of the growth rates w.r.t. increasing $d$ (see Fig.~\ref{sigma_vs_Pe_large_d}). Moreover, we plot the critical P\'eclet number for self-propulsion, $\Pe_c$, as a function of the drop-wall separation, $d$, in Fig.~\ref{sigma_vs_Pe_d_b}. The unbounded-case value $\Pe^{\infty}_c=4$ is recovered in the limit $d \to \infty$, and $\Pe_c$ reduces monotonically with $d$ and converges to $Pe_c^0 \approx 2$.  \nd{A power-law fit yields \nd{$\left( \Pe_c(d) - 2 \right) \sim d^{k_f}$, with $k_f \approx 0.46$} (see inset in Fig.~\ref{sigma_vs_Pe_d_b}), which suggests that $\Pe_c(d) - \Pe_c^0$ might scale as $\sim d^{1/2}$ for $d\rightarrow 0$.} This latter observation further confirms the emergence of a `lubrication regime' where the entire dynamics is dictated by the gap-scale transport.

One can use the reciprocal theorem to show that $V'_x,\;\Omega'_y \sim O(d^{-1/2})/\log(d)$ in such lubrication limit of $d \ll 1$, defining the auxiliary problems as the rigid motion of the sphere under the effect of a unit force along $\mathbf{e}_x$ and no torque (or unit torque along $\mathbf{e}_y$ and no force). Then, in the limit $d \ll 1$, $\left| \textbf{n} \cdot \hat{\mathbf{\sigma}} dS \right| \sim 1/\log(d)$ in the lubrication region~\cite{ONeill1967, Cooley1968, yariv2003} and for a constant rate of solute emission, Eq.~\eqref{AdvDiff_bcs}, $\left| \nabla_s c' \right|$ is $O(d^{-1/2})$~\cite{Yariv2016,Yariv2016b}. This means that the lubrication region has an $O(d^{-1/2})/\log (d)$ contribution to the integral on the left-hand-side of Eq.~\eqref{U_reciproc}, which is dominant over the (at most) $O(1)$ contribution from the outer region. Thus, for $d \ll 1$, $\int_{S} { \textbf{n} \cdot \boldsymbol{\hat\sigma} \cdot \nabla_s c'\;dS } \sim O(d^{-1/2})/ \log (d)$; which, combined with the unit forcing in the auxiliary problem yields the expected $O(d^{-1/2})/\log(d)$ scalings for both $V'_x$ and $\Omega'_y$. These large translational and rotational velocities are reminiscent of similarly large velocities obtained in studies of electrophoretic motion of a sphere moving parallel to a nearby wall (albeit with different scaling; see \cite{yariv2003}).

\section{Conclusion}\label{s:conclusion}

In this work, we analyzed the emergence of self-propulsion for an active drop hovering at a distance, $d$, from a rigid wall. 

To do so, we first characterised the properties of the axisymmetric levitating base state resulting from the balance between an external force, $\mathbf{F}^{ext}$, and a solute-polarity-induced hydrodynamic force, $\mathbf{F}_p = F_z \eb_z$, which maintains the drop stationary as it pumps fluid (see Fig.~\ref{probSchem}).

In the unbounded case ($d\rightarrow \infty$),  a symmetric bifurcation in the base state is observed at $\Pe=8$, from an isotropic state ($F_z = 0$), to a dipolar, asymmetric pumping state ($F_z > 0$). At higher $\Pe$, we showed the existence of a second bifurcation, from the \textit{dipolar, asymmetric} pumping state ($F_z > 0$), to a \textit{quadrupolar, symmetric} pumping state ($F_z = 0$). The presence of a wall (i.e. for moderate values of $d$) induces a polarization of the concentration field and a positive hydrodynamic force on the drop, $F_z > 0$, irrespective of $\Pe$. The dipolar--quadrupolar switch in the base state is still observed, but the symmetry about the drop's equator is lost. \nd{As $d$ is reduced further, the emergence of the quadrupolar mode is delayed to larger $\Pe$, until it is suppressed altogether for even smaller distances $d$, and the drop exclusively pumps fluid toward the wall.}

In a second step, we analyzed the linear stability of this base state to non-axisymmetric perturbations. For large $d$, the swimming speed of the drop is proportional to the first moment of the solute distribution at its surface so that a base state configuration that tends to enhance (resp. deplete) the solute concentration around the unbounded drop is expected to be more (resp. less) unstable. For finite $d$, the swimming speed is obtained as the average of the fluid slip on the drop surface weighted by the hydrodynamic traction in the corresponding rigid translation of the sphere. As $d$ is reduced, the wall induces both an increase in the concentration-induced phoretic slip near the drop's bottom pole, \textit{and} an increase of its influence on the drop's velocity (i.e. increased traction for the rigid body problem). The self-propulsion ability of the active drop is thus enhanced by the presence of the wall, and higher swimming velocities are thus expected. This is markedly different from the near-wall motion of other swimmers with \textit{a priori} prescribed surface slip (e.g. spherical squirmers), \nd{for which the strongest slip might not be localized around the most efficient region}, thus leading to increase or decrease of their swimming velocities due to confinement~\cite{Poddar2020}. \nd{It is also consistent with experiments wherein active drops have been observed to perform quasi-two-dimensional motion near a wall, while being pinned to it by gravity~\cite{Dwivedi2021}. An exploration of the drop's swimming speed as a function of its distance from the wall, however, has not been undertaken, to the best of our knowledge.}

In our analysis, the drop's response to concentration gradients was simply modelled as phoretic slip. Including Marangoni effects is only expected to change our results quantitatively, with the predominance of different base states and stability regimes (for fixed $d,\;\Pe$) being determined by the relative strengths of the phoretic and Marangoni effects. The key insights on the onset of self-propulsion along the wall (i.e. relative stability of dipolar and quadrupolar modes, wall effects on the onset of instability) are however expected to remain unchanged. \nd{It must be stressed however, that experimental relevance of our analysis to particles exhibiting a purely phoretic response is limited (e.g., a gradually
dissolving colloid). These colloids operate in the regime $\Pe \ll 1$~\cite{Moran2017,Boniface2019}; and not in the necessary condition for spontaneous propulsion, $\Pe \sim O(1)$, that is demonstrated here.}

A critical contribution to the destabilisation of the axisymmetric base state by the wall is the enhanced rigid-body traction (and thus enhanced slip influence) in the thin fluid gap separating the droplet and the wall as $d$ is reduced, which is intrinsically linked to the no-slip boundary condition at the wall. For a droplet swimming along a free-surface (i.e. with no-stress boundary condition), the picture may thus be fundamentally different as a reduction of the hydrodynamic stress is expected. The presence of stronger horizontal flows within the gap would further reduce the wall-induced concentration accumulation, thereby also reducing the phoretic slip and hydrodynamic force $F_z$. \nd{Such horizontal flows could also be present if the wall-solute interactions impart a phoretic mobility to the wall, say $\mathcal{M}_w$. The flows emerging due to $\mathcal{M}_w$ could promote or hamper solute accumulation near the wall (depending on the sign of $\mathcal{M}_w$), and thus affect self-propulsion non-trivially.} These observations suggest that confinement-induced destablisation of the drop may not be present (or as important) near a free-surface \nd{and/or a surface with instrinsic mobility}, but this question should be analysed further and represents a natural extension of the present work.

The approach and framework of the present analysis could also easily be generalised to analyse the  stability of axisymmetric bound states of interacting active drops as observed~\cite{Lippera2020b}, and thus provide some insight on their occurence (or lack thereof) and the collective dynamics of active droplets in suspensions~\cite{Thutupalli2011, Kruger2016}.

A linear stability analysis reveals important features and characteristics of the onset of self-propulsion and the swimming mechanism near the threshold. Analysing in detail the steady near-wall motion however requires solving for the full non-linear problem, e.g. with an extension of the framework of Ref.~\cite{Lippera2020} to non-axisymmetric fields. Such generalisation is however no easy task as the non-linearity of the advection-diffusion equation indeed couples the different azimuthal contributions to the concentration field~\cite{Morozov2019b}, as well as the wall-normal and wall-parallel components of the drop's motion. Such coupling however opens the possibility for intriguing wall-mediated drop dynamics (e.g., existence of limit cycles, fixed points). A comparison of these behaviors against equivalent theoretical analyses of squirmers~\cite{Ishikawa2006, Ishimoto2013} and Janus particles~\cite{SharifiMood2016} would further help towards uniquely identifying the role of chemo-dynamics and solute advection on interactions of active drops with nearby surfaces, or with each other.

\begin{acknowledgments}
This work was supported by the European Research Council (ERC) under the European Union’s Horizon 2020 research and innovation program (Grant Agreement No. 714027 to S.M.).
\end{acknowledgments}

\appendix

\section{Asymptotic solution for the force experienced by the drop in the equilibrium position}\label{app_asymp}

\begin{figure}[t]
\begin{center}
      \includegraphics[width=0.75\linewidth]{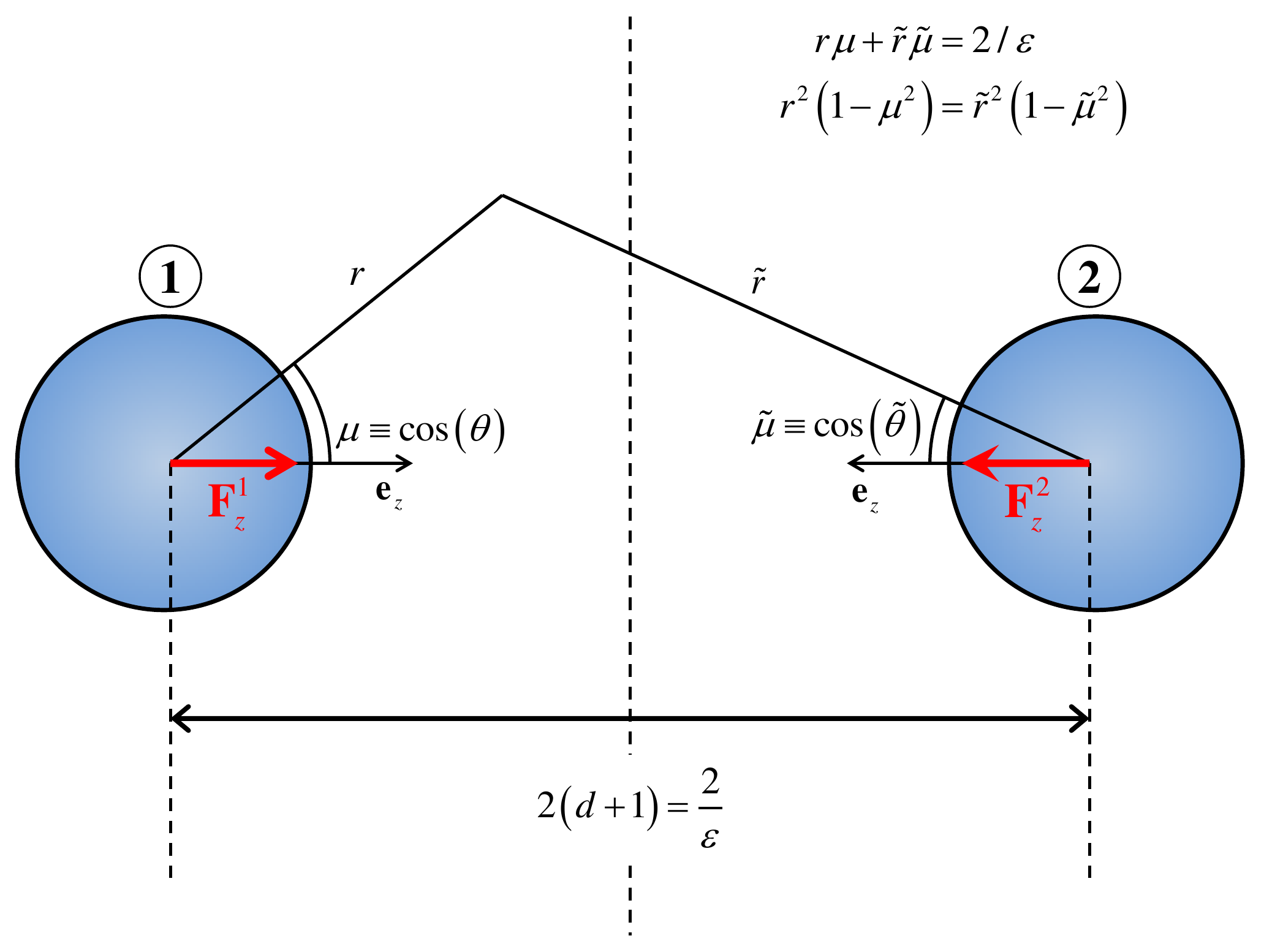}
  \caption{Coordinate systems and notations used in the asymptotic analysis for evaluation of the force experienced by a \textit{fixed}, phoretic particle-1 in the presence of an identical phoretic particle-2, placed `far away' at a dimensionless center-to-center distance $2(d+1)$. We show in Appendix \ref{app_asymp} that the leading order force experienced by each particle in this configuration is identical to that experienced if the other particle were replaced by a wall placed along the mid-plane.}
  \label{probSchem_asymp}
\end{center}
\end{figure}

We consider the interaction between two fixed drops separated by a center-to-center distance of $2\left( d+1 \right)$, and define $\epsilon = 1/\left( d+1 \right)$ (Fig. \ref{probSchem_asymp}). We show that, to leading order when $\epsilon \ll 1$, there is no difference between this system and the one which is the focus of our study (i.e., an active drop near a rigid, passive wall). In what follows, we will show that leading order contribution of the wall to the isotropic solute distribution is $O \left( \epsilon \right)$, while that to the fluid flow is $O \left( \epsilon^2 \right)$. 

In the absence of the second particle/drop, and for $\Pe < 8$, the solute distribution around particle 1 is isotropic,
\begin{equation}\label{c_iso}
    c_{iso}(r) = \frac{1}{r},
\end{equation}
and hence the fluid is quiescent ($\nabla_sc_{iso} = \textbf{0}$). When $\Pe=0$, the solute transport problem is linear and the presence of the second particle can be accounted for using successive reflections~\cite{Varma2019, Rallabandi2019}. Here, we show that as long as $\Pe<8$, we can still use a similar method to determine the interaction between two distant, fixed phoretic particles. We begin by writing the isotropic solute distribution around the second particle, in the coordinate system of the first particle:
\begin{equation}\label{c_tilde}
    \tilde{c}_{iso}(\tilde{r}) = \frac{1}{\tilde{r}} = \frac{\epsilon}{2} + \epsilon^2\frac{\mu r}{4} + O(\epsilon^3).
\end{equation}
The `combined' solute concentration, $c(r)$, due to two active particles fixed at a center-to-center separation $2/\epsilon = 2(d+1)$ (with, $d\gg1$) is thus given by:
\begin{align}\label{c_combined1}
    c(r,\mu) & = \frac{1}{r} + \frac{1}{\tilde{r}}+ c'(r,\mu), \nonumber \\
         & = \frac{1}{r} + \frac{\epsilon}{2} + \epsilon^2\frac{\mu r}{4} + c'(r,\mu)+O(\epsilon^3),
\end{align}
where, $c'(r,\mu)$ is a correction that incorporates the interaction (chemical and/or hydrodynamic) between the two particles. At this stage, we do not know the order in $\epsilon$ at which this correction becomes important, but it is expected to be less than $O(1)$ as the isotropic concentration should be recovered for $\epsilon\rightarrow 0$. Substituting Eq.~\eqref{c_combined1} in the steady-state advection-diffusion equation, the boundary conditions for $c$ and $\textbf{u}$ at the drop surface, we obtain:
\begin{align}\label{adv_diff1}
    Pe \epsilon^2 \textbf{u} \cdot \nabla \left( \frac{r \mu}{4} \right) + Pe \textbf{u} \cdot \nabla \left( \frac{1}{r} + c' \right) & = \nabla^2 c',
\end{align}
\begin{align}\label{nGradC1}
\left. \frac{\partial}{\partial r} \left( \epsilon^2\frac{\mu r}{4} + c' \right) \right|_{r=1} & = 0,
\end{align}
\begin{align}\label{grad_sC1}
    \left. \textbf{u} \right|_{r=1} & = \left.-\frac{\sqrt{1-\mu^2}}{r} \frac{\partial}{\partial \mu} \left( \epsilon^2\frac{\mu r}{4} + c' \right) \right|_{r=1}.
\end{align}
Clearly, the $O(\epsilon^2)$ balance in Eq.~\eqref{nGradC1} can only be satisfied if the correction $c'$ itself is $O(\epsilon^2)$, thus we define,
\begin{equation}\label{c_2}
c'(r,\mu) = \epsilon^2 c_2(r,\mu).
\end{equation}
Eq.~\eqref{grad_sC1} then shows that the leading order fluid flow due to presence of the second particle is $O(\epsilon^2)$. Therefore, the correction due to the advection term, $\textbf{u} \cdot \nabla \tilde{c}_{iso}(r)$, is $O(\epsilon^4)$, whereas that due to the diffusion term, $\nabla^2 \tilde{c}_{iso}(r)$, is $O(\epsilon^3)$. Thus, any error introduced into our analysis due the non-linear advective term is an order of magnitude smaller than the diffusion term. This allows us to proceed toward solving for the leading order flow induced around particle-1 due to the chemical field of particle-2. This flow-field can be written as,
\begin{equation}\label{u_lead}
    \textbf{u}\left( r, \mu \right) = \epsilon^2 \textbf{u}_2\left( r, \mu \right) + O(\epsilon^3).
\end{equation}
After using Eqs.~\eqref{c_2} and \eqref{u_lead} in Eqs.~\eqref{adv_diff1} to \eqref{grad_sC1} and collecting the resulting $O(\epsilon^2)$ terms, we obtain:
\begin{equation}\label{advdiff_gde_e2}
    Pe \textbf{u}_2 \cdot \nabla (1/r) = \nabla^2 c_2,
\end{equation}
subject to,
\begin{equation}\label{advdiff_bc_e2}
    \left. \frac{\partial c_2}{\partial r} \right|_{r=1} = -\frac{\mu}{4},
\end{equation}
and,
\begin{align}\label{stokes_bcs_e2}
    \left. \textbf{u}_2 \cdot \textbf{e}_r \right|_{r=1} & = 0, \nonumber \\
    \left. \textbf{u}_2 \cdot \textbf{e}_{\theta} \right|_{r=1} & = -\sqrt{1-\mu^2} \left( \frac{1}{4} + \frac{\partial c_2}{\partial \mu} \right).
\end{align}
where, $\textbf{u}_2$ is, of course, governed by the Stokes equation,
\begin{equation}\label{stokes_gde_e2}
    -\nabla p_2 + \nabla^2 \textbf{u}_2= \textbf{0},
\end{equation}
and subject to far-field decay, i.e.,
\begin{equation}\label{stokes_far_e2}
    \left| \textbf{u}_2 \right| \to 0,
\end{equation}
as $r \to \infty$. Equations.~\eqref{advdiff_gde_e2} to \eqref{stokes_far_e2} can be solved using spherical harmonic expansions for $\textbf{u}_2\left( r,\mu \right)$ and $c_2\left( r,\mu \right)$~\cite{Michelin2013}, to obtain:
\begin{equation}\label{u2}
    \epsilon^2 \textbf{u}_2\left( r, \mu \right) = \frac{3\epsilon^2\mu}{(Pe-8)}\left( \frac{1}{r^3} - \frac{1}{r} \right)\textbf{e}_r + \frac{3\epsilon^2\sqrt{1-\mu^2}}{2(Pe-8)}\left( \frac{1}{r^3} + \frac{1}{r} \right)\textbf{e}_{\mu},
\end{equation}
and,
\begin{equation}\label{c2}
    c_2\left( r, \mu \right) = \mu \left( \frac{r}{4} + \frac{2Pe-1}{Pe-8}\frac{1}{r^2} - \frac{3Pe}{Pe-8}\frac{2r^2+1}{4r^3} \right).
\end{equation}

Note that the interaction represented via Eqs.~\eqref{u2} and \eqref{c2} is purely chemical, i.e., we have obtained corrections to the isotropic state around particle-1 due to \textit{only} the chemical field of particle-2. The hydrodynamic interactions would entail a calculation of the concentration and flow-field corrections due to the flow $\epsilon^2 \textbf{u}_2$, which would correspond to an $O(\epsilon^3)$ correction, as $\textbf{u}_2 \sim \epsilon$ for $r \gg 1$. We can thus conclude that the present analysis would remain unchanged if we were to replace particle-2 with a wall along the plane of symmetry. In the latter case, the only effect of the wall would be to force the no-flux condition for the solute concentration,
\begin{equation}
\left. \textbf{n} \cdot \nabla c \right|_{\mathscr{W}} = 0,    
\end{equation}
which is already incorporated in the expansion Eq.~\eqref{c_combined1}. The rest of the analysis remains the same. The velocity boundary condition at the wall would enter the analysis at $O(\epsilon^3)$, in the form of a reflection of an $O(\epsilon^2)$ Stokeslet flow. The solute concentration is therefore given by,
\begin{equation}\label{c_corr_far-field}
    c\left(r,\mu\right) = \frac{1}{r} + \frac{\epsilon}{2} + \epsilon^2 \left\{ \frac{r\mu}{4} + c_2\left(r,\mu\right) \right\} + O(\epsilon^3),
\end{equation}
from which we can calculate the force via the surface integration \cite{Yariv2016}:
\begin{equation}\label{Fz_ana}
    F^1_z = \frac{3}{2} \textbf{e}_z \cdot \int_{S} { \nabla_sc\;dS } = \frac{12 \pi}{\left( d+1 \right)^2 \left( Pe-8 \right)}.
\end{equation}

\section{Projection of the axisymmetric problem (base state)}

\subsection{Projection of the advection-diffusion equation}

The tensorial expressions given in this Section are exactly the same as those in the Appendix-A.1. of \cite{Lippera2020}, but we repeat these here for the sake of completeness. The tensors $\textbf{B}^i$, $\textbf{H}^1$ and $\textbf{A}^i$ in Eq.~\eqref{AdvDiff_axiSymm_disc} are:

\begin{align}\label{B_tens_axisymm}
  B_{pnk}^{1} & = \frac{1}{{{a}^{3}}}\left( \frac{3\sinh \left( \lambda \xi  \right)}{2}S_{knp}^{0}-\frac{k\left( k+1 \right)}{2}\sinh \left( \lambda \xi  \right)Q_{knp}^{0} \right), \nonumber \\ 
 B_{pnk}^{2} & = \frac{1}{\lambda {{a}^{3}}}\left( -\cosh \left( \lambda \xi  \right)S_{knp}^{0}+S_{knp}^{1}+\frac{1}{2}R_{nkp}^{0} \right), \nonumber \\ 
 B_{pnk}^{3} & = \frac{1}{\lambda {{a}^{3}}}\left\{ \frac{3}{2}R_{nkp}^{0}-k\left( k+1 \right)\left( \cosh \left( \lambda \xi  \right)Q_{knp}^{0}-Q_{knp}^{1} \right) \right\}.  
\end{align}

\begin{equation}\label{H_tens_axisymm}
    {H^1_{pn}}=\sqrt{2}\sum\limits_{k=0}^{\infty }{Q_{knp}^{0}{{\text{e}}^{-\left( k+1/2 \right)\left| \lambda \xi  \right|}}}.
\end{equation}

\begin{align}\label{A_tens_axisymm}
 A_{pn}^{2} & = \frac{\sqrt{2}}{{{\lambda }^{2}}{{a}^{2}}}\sum\limits_{k=0}^{\infty }{\left\{ {{\cosh }^{2}}\left( \lambda \xi  \right)Q_{knp}^{0}-2\cosh \left( \lambda \xi  \right)Q_{knp}^{1}+Q_{knp}^{2} \right\}{\textrm{e}^{-\left( k+1/2 \right)\left| \lambda \xi  \right|}}}, \nonumber \\ 
 A_{pn}^{1} & = -{{\lambda }^{2}}{{\left( n+\frac{1}{2} \right)}^{2}}A_{pn}^{2}.  
\end{align}
In Eqs.~\eqref{B_tens_axisymm} to \eqref{A_tens_axisymm}, $Q^i_{knp}$, $S^i_{knp}$ and $R^i_{knp}$ are the following integrals involving the zeroth-order Legendre polynomials and their derivatives:
\begin{equation}\label{Q_int}
 Q_{knp}^{i} = \int\limits_{-1}^{1}{{{\mu }^{i}}{{L}_{k}}{{L}_{n}}{{L}_{p}}d\mu },
\end{equation}
\begin{equation}\label{S_int}
  S_{knp}^{i} = \int\limits_{-1}^{1}{{{\mu }^{i}}\left( 1-{{\mu }^{2}} \right){{{{L}'}}_{k}}{{{{L}'}}_{n}}{{L}_{p}}d\mu },
\end{equation}
\begin{equation}\label{R_int}
  R_{knp}^{i} = \int\limits_{-1}^{1}{{{\mu }^{i}}\left( 1-{{\mu }^{2}} \right){{L}_{k}}{{{{L}'}}_{n}}{{L}_{p}}d\mu }.
\end{equation}

\subsection{Projection of the hydrodynamic boundary conditions}

We first describe the projection of the tangential component of the boundary condition, Eq.~\eqref{Stokes_bcs_drop} (or, Eq.~\eqref{u_tan_drop}), on the drop surface. From the expansion for $c$, Eq.~\eqref{c_axisymm}, we can evaluate $\partial c/\partial \mu$,
\begin{align}\label{del_c_del_mu}
  \frac{\partial c}{\partial \mu} = \Gamma^{1/2} \sum\limits_{n=0}^{\infty}{c_n\left( \xi, t \right) L'_n \left( \mu \right)} - \frac{1}{2 \Gamma^{1/2}} \sum\limits_{n=0}^{\infty}{c_n\left( \xi, t \right) L_n \left( \mu \right)}.
\end{align}
Eq.~\eqref{u_tan_drop} can then be projected onto $\sqrt{1-\mu^2}L'_p(\mu)$ to obtain:
\begin{align}\label{u_base_tan_proj_drop}
    & \sum\limits_{n=1}^{N}{ \left. \left[ \frac{S^1_{pn0} - S^0_{pn0} \cosh(\lambda)}{\lambda a}  \frac{\partial U_n}{\partial \xi} \right] \right|_{\xi=1}} + \frac{3 \sinh (\lambda)}{2a} \frac{2p(p+1)}{2p+1}U_p(\xi=1,t) \nonumber \\
    & = \sum\limits_{n=0}^{N}{\left. c_n(\xi=1,t) \left[ S^0_{pn0}\cosh^2(\lambda) - 2S^1_{pn0}\cosh(\lambda) + S^2_{pn0} + \frac{R^1_{np0}}{2} - \frac{R^0_{np0} \cosh(\lambda)}{2} \right] \right|_{\xi=1}}.
\end{align}
The vanishing normal velocity at the drop surface simply yields,
\begin{equation}\label{u_base_norm_proj_drop}
    \left. U_n \right|_{\xi=1} = 0.
\end{equation}
Similarly, the conditions for vanishing normal and tangential velocities at the wall yield, respectively,
\begin{equation}\label{u_base_norm_proj_wall}
    \left. U_n \right|_{\xi=0} = 0,
\end{equation}
and,
\begin{equation}\label{u_base_tan_proj_wall}
    \left. \frac{\partial U_n}{\partial \xi} \right|_{\xi=0} = 0.
\end{equation}
Eqs.~\eqref{u_base_tan_proj_drop} to \eqref{u_base_tan_proj_wall} along with the definition of $U_n(\xi, t)$, Eq.~\eqref{Un}, allow us to obtain the coefficients $\left[ \alpha_n, \beta_n, \gamma_n, \delta_n \right]$, and express the base state flow, in terms of the surface concentration modes $c_n(\xi=1,\mu,t)$ at each instant.

\section{Projection of the hydrodynamic boundary conditions in the non-axisymmetric case}

The hydrodynamics problem in the general (non-axisymmetric) case involves $7N+2$ unknowns, i.e., $\mathcal{U}_H$, $V'_x$ and $\Omega'_y$. The vector $\mathcal{U}_H$ contains the following $7N$ coefficients from the velocity field expansion in Eqs.~\eqref{uprime_z} to \eqref{gamma_m_chi_m}:
\begin{equation}\label{nonAxisymm_unknowns_hydro}
 \mathcal{U}_H \equiv \left[A^1_1 ... A^1_N,\;B^1_1 ... B^1_N,\;C^1_1 ... C^1_N,\;E^1_2 ... E^1_N,\;F^1_2 ... F^1_N,\;G^1_0 ... G^1_N,\;H^1_0 ... H^1_N \right].
\end{equation}
The $7N+2$ linear equations required to solve for these hydrodynamic variables are obtained as follows:
\renewcommand{\labelenumi}{\arabic{enumi}}
\begin{enumerate}[(i)]
    \item The projection of the continuity equation and the wall boundary condition, Eq.~\eqref{BC_wall_nonAxisymm}, onto $P^1_k\left( \mu \right) \cos\left( \phi \right)$ yields $4N$ equations. The linear equations corresponding to these projections are given in \cite{Mozaffari2016} (see their equations B15, B16, B28 and B29 for $m=1$), and we avoid writing them here for the sake of brevity.
    \item The integrals involved in the calculation of $F_x$ and $T_y$ (Eqs.~\eqref{force_int}) show that these depend only on the $m=1$ velocity coefficients, $\mathcal{U}_H$. The force- and torque-free conditions thus yield two more equations (see Eqs.~\eqref{force_disc} and \eqref{torque_disc}).
    \item The projection of the $\left( \textbf{e}_{z}, \textbf{e}_{\rho}, \textbf{e}_{\phi} \right)$ components of the boundary condition on the drop surface, Eq.~\eqref{BC_drop_nonAxisymm}, onto $P^1_k\left( \mu \right) \cos\left( \phi \right)$ yields $3N$ linear equations relating the coefficients $\mathcal{U}_H$, $V'_x$, $\Omega'_y$, and the surface concentration modes $\textbf{C}'_{surf} \equiv \left[ c^1_1(1),\;c^1_2(1),...,c^1_N(1) \right]$. These systems of equations are written explicitly in Eqs.~\eqref{ur_uphi_BC_short} to \eqref{uz_BC}.
\end{enumerate}

In this way, the $7N+2$ unknowns of the hydrodynamic problem are completely defined in terms of the surface concentration modes, and the hydrodynamics problem is formally solved, i.e., $\textbf{u}'$ is known as a linear function of $\textbf{C}'_{surf}$. Thus, a solution of the eigenvalue problem defined by the advection-diffusion Eq.~\eqref{gde_stab}, for the unknown functions $\mathbf{C}'$ and the growth rate $\sigma$, completes our analysis.

\subsection{Force- and torque-free conditions}

In the general bi-spherical coordinate system used here, the hydrodynamic force on the drop depends only on the coefficients $A^1_n$, $B^1_n$, $G^1_n$ and $H^1_n$ \cite{Lee1980, Mozaffari2016}, as:
\begin{equation}\label{force_disc}
    F_x = -\sqrt{2} \pi a \sum_{n=0}^{\infty}{ \left\{ (G^1_n+H^1_n) - n(n+1)(A^1_n+B^1_n) \right\} } = 0;
\end{equation}
while the hydrodynamic torque depends only on the coefficients $A^1_n$, $B^1_n$, $C^1_n$, $G^1_n$ and $H^1_n$, as:
\begin{align}\label{torque_disc}
    T_y = & -\sqrt{2} \pi a^2 \sum_{n=0}^{\infty}{ n(n+1)\left\{ 2C^1_n + \coth(\lambda)(A^1_n+B^1_n) \right\} } \nonumber \\
    & -\sqrt{2} \pi a^2 \sum_{n=0}^{\infty}{\left\{ 2n + 1 - \coth(\lambda) \right\}(G^1_n+H^1_n)} = 0.
\end{align}

\subsection{Projection of the phoretic boundary condition}

The linear equations corresponding to the boundary conditions at the wall and the continuity equation are given in \cite{Mozaffari2016} (see their equations B15, B16, B28 and B29 for $m=1$), and we avoid writing them here for the sake of brevity. We focus here on the linear equations arising from the projections of Eq.~\eqref{BC_drop_nonAxisymm}. We first eliminate the $\phi$ dependence by projecting the $u'_r$ and $u'_{\phi}$ boundary conditions onto $\cos(\phi)$ and $\sin(\phi)$, respectively. Thereafter, we can add (resp. subtract) the ensuing equations and project them onto $P^2_l(\mu)$ (resp. $P^0_l(\mu)$), to obtain the following:
\begin{align}\label{ur_uphi_BC_short}
  & -\frac{1}{2\left( 2l-1 \right)}\left[ A_{l-1}^{1}\sinh \left\{ \left( l-1/2 \right)\lambda  \right\}+B_{l-1}^{1}\cosh \left\{ \left( l-1/2 \right)\lambda  \right\} \right] \nonumber \\ 
 & +\frac{1}{2\left( 2l+3 \right)}\left[ A_{l+1}^{1}\sinh \left\{ \left( l+3/2 \right)\lambda  \right\}+B_{l+1}^{1}\cosh \left\{ \left( l+3/2 \right)\lambda  \right\} \right] \nonumber \\ 
 & -\frac{\left( l-2 \right)}{\left( 2l-1 \right)}\left[ E_{l-1}^{1}\sinh \left\{ \left( l-1/2 \right)\lambda  \right\}+F_{l-1}^{1}\cosh \left\{ \left( l-1/2 \right)\lambda  \right\} \right] \nonumber \\ 
 & +\cosh \left( \lambda  \right)\left[ E_{l}^{1}\sin \left\{ \left( l+1/2 \right)\lambda  \right\}+F_{l}^{1}\cos \left\{ \left( l+1/2 \right)\lambda  \right\} \right] \nonumber \\ 
 & -\frac{\left( l+3 \right)}{\left( 2l+3 \right)}\left[ E_{l+1}^{1}\sin \left\{ \left( l+3/2 \right)\lambda  \right\}+F_{l+1}^{1}\cos \left\{ \left( l+3/2 \right)\lambda  \right\} \right] = \nonumber \\ 
 & \frac{\left( 2l+1 \right)\left( l-2 \right)!}{2\left( l+2 \right)!}\sum\limits_{n=1}^{\infty }{\left\{ \Theta _{ln}^{2}+\zeta _{ln}^{2}+\nu _{ln}^{2}+\kappa _{ln}^{2} \right\}{{c}^1_n}\left( \xi =1 \right)},
\end{align}
and,
\begin{align}\label{ur_uphi_BC_long}
    & \frac{l\left( l-1 \right)}{2\left( 2l-1 \right)}\left[ A_{l-1}^{1}\sinh \left\{ \left( l-1/2 \right)\lambda  \right\}+B_{l-1}^{1}\cosh \left\{ \left( l-1/2 \right)\lambda  \right\} \right] \nonumber \\ 
 & -\frac{\left( l+1 \right)\left( l+2 \right)}{2\left( 2l+3 \right)}\left[ A_{l+1}^{1}\sinh \left\{ \left( l+3/2 \right)\lambda  \right\}+B_{l+1}^{1}\cosh \left\{ \left( l+3/2 \right)\lambda  \right\} \right] \nonumber \\ 
 & -\frac{l}{\left( 2l-1 \right)}\left[ G_{l-1}^{1}\sinh \left\{ \left( l-1/2 \right)\lambda  \right\}+H_{l-1}^{1}\cosh \left\{ \left( l-1/2 \right)\lambda  \right\} \right] \nonumber \\ 
 & +\cosh \left( \lambda  \right)\left[ G_{l}^{1}\sin \left\{ \left( l+1/2 \right)\lambda  \right\}+H_{l}^{1}\cos \left\{ \left( l+1/2 \right)\lambda  \right\} \right] \nonumber \\ 
 & -\frac{\left( l+1 \right)}{\left( 2l+3 \right)}\left[ G_{l+1}^{1}\sin \left\{ \left( l+3/2 \right)\lambda  \right\}+H_{l+1}^{1}\cos \left\{ \left( l+3/2 \right)\lambda  \right\} \right] = \nonumber \\ 
 & \frac{\left( 2l+1 \right)}{2}\sum\limits_{n=1}^{\infty }{\left\{ \Theta _{ln}^{0}+\zeta _{ln}^{0}-\nu _{ln}^{0}+\kappa _{ln}^{0} \right\}{{c}^1_n}\left( \xi =1 \right)} \nonumber \\ & + \left( 2l+1 \right) \int\limits_{-1}^{1}{ \sqrt{\cosh \left( \lambda  \right)-\mu }\left\{ {V'_x}-{\Omega'_y}\frac{1-\mu \cosh \left( \lambda  \right)}{\cosh \left( \lambda  \right)-\mu } \right\}P_{l}^{0}\left( \mu  \right)d\mu }.
\end{align}
In Eq.~\eqref{ur_uphi_BC_short} we have $2 \le l \le N$, whereas in Eq.~\eqref{ur_uphi_BC_long} we have $0 \le l \le N$. The tensors $\left\{ \Theta^j_{ln},\; \zeta^j_{ln},\; \kappa^j_{ln},\; \nu^j_{ln} \right\}$, with $j=0,2$, are integrals involving: (i) the appropriate Legendre polynomials, and, (ii) the dot products of the basis vectors in cylindrical coordinates with those of bi-spherical coordinates, Eq.~\eqref{erho_ez_exi_emu} (see Eqs.~\eqref{Theta_zeta_nu_tens} and \eqref{kappa_tens}). The projection of the $u'_z$ boundary condition yields:

\begin{align}\label{uz_BC}
  & \frac{\sinh \left( \lambda  \right)}{2}\left[ A_{l}^{1}\sinh \left\{ \left( l+1/2 \right)\lambda  \right\}+B_{l}^{1}\cosh \left\{ \left( l+1/2 \right)\lambda  \right\} \right] \nonumber \\ 
 & -\frac{\left( l-1 \right)}{\left( 2l-1 \right)}C_{l-1}^{1}\sinh \left\{ \left( l-1/2 \right)\lambda  \right\} \nonumber \\
 & +C_{l}^{1}\cosh \left( \lambda  \right)\sinh \left\{ \left( l+1/2 \right)\lambda  \right\} \nonumber \\
 & -\frac{\left( l+2 \right)}{\left( 2l+3 \right)}C_{l+1}^{1}\sinh \left\{ \left( l+3/2 \right)\lambda  \right\} \nonumber \\
 & = \frac{\sinh (\lambda)}{2a} \left[ c^1_{n-2}(1)\frac{(l-1)(l-2)}{(2l-1)} - c^1_{n-1}(1)\frac{2 \cosh (\lambda) (l-1)^2}{(2l-1)} \right] \nonumber \\ 
 & + \frac{\sinh (\lambda)}{a} c^1_{n}(1) \frac{(-2l^2 - 2l + 3)}{(2l-1)(2l+3)} \nonumber \\
 & + \frac{\sinh (\lambda)}{2a} \left[ c^1_{n+1}(1)\frac{2 \cosh(\lambda) (l+2)^2}{(2l+3)} - c^1_{n+2}(1)\frac{(l+2)(l+3)}{(2l+3)} \right] \nonumber \\
 & - \Omega'_y \times \frac{2l+1}{2l(l+1)} \int\limits_{-1}^{1}{ \sqrt{ \frac{1 - \mu^2}{\cosh \left( \lambda  \right)-\mu} } \sinh(\lambda) P_{l}^{1}\left( \mu  \right)d\mu },
\end{align}
where $l \ge 1$. We next provide the integrals that define the tensors $\left\{ \Theta^j_{ln},\; \zeta^j_{ln},\; \kappa^j_{ln},\; \nu^j_{ln} \right\}$ in Eqs.~\eqref{ur_uphi_BC_short} and \eqref{ur_uphi_BC_long}, with $j=0,2$.
\begin{align}\label{Theta_zeta_nu_tens}
   \Theta _{ln}^{j} &= \int\limits_{-1}^{1}{\frac{\mu \cosh \left( \lambda  \right)-1}{2a}\sqrt{1-{{\mu }^{2}}}P_{l}^{j}\left( \mu  \right)P_{n}^{1}\left( \mu  \right)d\mu }, \nonumber \\ 
   \zeta _{ln}^{j} &= \int\limits_{-1}^{1}{ \left[ \frac{\cosh \left( \lambda  \right)-\mu }{2a}\left\{ 1-\mu \cosh \left( \lambda  \right) \right\}P_{l}^{j}\left( \mu  \right)\times n\left( n+1 \right)P_{n}^{0}\left( \mu  \right) \right] d\mu }, \\ 
   \nu _{ln}^{j} &= \int\limits_{-1}^{1}{\frac{{{\left\{ \cosh \left( \lambda  \right)-\mu  \right\}}^{2}}}{a\sqrt{1-{{\mu }^{2}}}}P_{l}^{j}\left( \mu  \right)P_{n}^{1}\left( \mu  \right)d\mu }. \nonumber
\end{align}

\begin{equation}\label{kappa_tens}
    \kappa _{ln}^{j}=\left\{ \begin{matrix}
   0,\text{ }n=1  \\
   \int\limits_{-1}^{1}{\frac{\cosh \left( \lambda  \right)-\mu }{2a}\left\{ \mu \cosh \left( \lambda  \right)-1 \right\}P_{l}^{j}\left( \mu  \right)P_{n}^{2}\left( \mu  \right)d\mu },\text{ }n\ge 2  \\
\end{matrix} \right. .
\end{equation}

\section{Projection of the linearised non-axisymmetric advection-diffusion equation}

In this section, we detail the different terms emerging from the projection of Eq.~\eqref{gde_stab} onto the modes $P^1_k(\mu) \cos (\phi)$. As mentioned in Section \ref{s:stability2}, we replace the general expressions for the various fields, Eqs.~\eqref{uprime_z} to \eqref{c_nonAxisymm}, and the base state solution from Section \ref{s:baseState} into Eq.~\eqref{gde_stab}, and then divide the entire equation by $\Gamma$ before evaluating the projections.

\subsection{Time-derivative of the concentration field}

The projection of the $\sigma c'$ term, evaluated at the position $\xi = \xi_i$, yields:
\begin{equation}\label{disc_sigma_cprime}
    \left. \left< \sigma c' \right> \right|_{\xi=\xi_i}  = \mathbf{H}^2 \cdot \mathbf{C}' = \sigma \sum\limits_{n=1}^{\infty}{ H_{kn}^{2}\left( \xi_i \right)c_{n}^{1}\left( \xi_i \right) },
\end{equation}
where, the $\left< ... \right>$ denotes the projection operation:
\begin{equation}\label{genProj}
    \left\langle f\left( {{\xi }_{i}},\mu ,\phi  \right) \right\rangle =\int\limits_{-1}^{1}{\left\{ \int\limits_{0}^{2\pi }{f\left( {{\xi }_{i}},\mu ,\phi  \right)\cos \left( \phi  \right)d\phi } \right\}P_{k}^{1}\left( \mu  \right)d\mu }.
\end{equation}
The tensor $H^2_{kn}$ is given by:
\begin{equation}\label{T_m_kn}
    H^2_{kn}(\xi_i)=\sqrt{2}\sum\limits_{p=0}^{\infty }{ \left[ \int\limits_{-1}^{1} {\text{e}}^{-\left( p+1/2 \right)\left| \lambda \xi  \right|} L_p\left( \mu  \right){P_{k}^{1}\left( \mu  \right)P_{n}^{1}\left( \mu  \right) d\mu } \right] },
\end{equation}
where, $L_p(\mu) \equiv P^0_p(\mu)$ is the associated Legendre polynomial of zeroth order and $p$-th degree.

\subsection{Advection of the solute perturbation by the base flow}

The projection of the $\textbf{u}^b \cdot \nabla c'$ term at position $\xi_i$ onto $P^1_k(\mu) \cos (\phi)$, is given by:
\begin{align}\label{disc_ub_grad_cprime}
\left. \left< \mathbf{u}^b\cdot \nabla {{c}'} \right> \right|_{\xi=\xi_i} & = \mathbf{B}^4 \cdot \mathbf{C}' + \mathbf{B}^5 \cdot \frac{d \mathbf{C}'}{d \xi }, \nonumber \\ & =
\sum\limits_{n=1}^{\infty }{B_{kn}^{4}\left( {{\xi }_{i}} \right){{{{c}}}^1_n}\left( {{\xi }_{i}} \right)} + \sum\limits_{n=1}^{\infty }{B_{kn}^{5}\left( {{\xi }} \right)\left.\frac{d c^1_n}{d \xi}\right|_{\xi=\xi_i}},
\end{align}
where the $B_{kn}^4\left( \xi_i \right)$ and $B_{kn}^5\left( \xi_i \right)$ are integrals involving the $\xi$ and $\mu$ components of $\textbf{u}^b$:
\begin{equation}\label{sc_C_zero_kn}
    B_{kn}^{4}\left( {{\xi }_{i}} \right)=\int\limits_{-1}^{1}{ P_{k}^{1}\left( \mu  \right) \left[ u_{\xi }^{b}\left( {{\xi }_{i}},\mu  \right)F_{n}^{1,2}\left( {{\xi }_{i}},\mu  \right)+u_{\mu }^{b}\left( {{\xi }_{i}},\mu  \right)\left\{ F_{n}^{1,3}\left( {{\xi }_{i}},\mu  \right)+F_{n}^{1,4}\left( {{\xi }_{i}},\mu  \right) \right\} \right] d\mu },
\end{equation}
and,
\begin{equation}\label{sc_C_plus_kn}
    B_{kn}^{5}\left( {{\xi }_{i}} \right) = \int\limits_{-1}^{1}{u_{\xi }^{b}\left( {{\xi }_{i}},\mu  \right) P_{k}^{1}\left( \mu  \right) F_{n}^{1,1}\left( {{\xi }_{i}},\mu  \right)d\mu }.
\end{equation}
The functions $F^{1,1}_{n} \left( \xi_i, \mu \right) $ to $F^{1,4}_{n} \left( \xi_i, \mu \right) $ are:

\begin{equation}\label{F1_1n}
    F^{1,1}_n(\xi,\mu) = \frac{\Gamma^{1/2}}{a \lambda}P^1_n(\mu),
\end{equation}

\begin{equation}\label{F2_1n}
    F^{1,2}_n(\xi,\mu) = \frac{\Gamma^{-1/2}}{2 a} \sinh(\lambda \xi) P^1_n(\mu),
\end{equation}

\begin{equation}\label{F3_1n}
    F^{1,3}_n(\xi,\mu) = \frac{\Gamma^{1/2}}{a} \frac{dP^1_n}{d \mu} \sqrt{1 - \mu^2},
\end{equation}

\begin{equation}\label{F4_1n}
    F^{1,4}_n(\xi,\mu) = -\frac{\Gamma^{-1/2}}{2a} \sqrt{1 - \mu^2} P^1_n(\mu).
\end{equation}

\subsection{Advection of the base state solute distribution by the perturbation flow}
The $\textbf{u}' \cdot \nabla c^b$ term, when evaluated at $\xi = \xi_i$ and then projected onto the mode $P^1_k(\mu) \cos (\phi)$ will yield:
\begin{align}\label{disc_uprime_grad_cb}
  \left. \left< \mathbf{{u}'}\cdot \nabla {{c}^b} \right> \right|_{\xi=\xi_i} & = \mathbf{B}^6 \cdot \mathcal{U}_H
  \nonumber \\ & = \sum\limits_{n=1}^{\infty }{\left\{ \mathcal{A}_{kn}^{1}\left( \xi_i \right)A_{n}^{1}+\mathcal{B}_{kn}^{1}\left( \xi_i \right)B_{n}^{1}+\mathcal{C}_{kn}^{1}\left( \xi_i \right)C_{n}^{1} \right\}} \nonumber \\ & +\sum\limits_{n=2}^{\infty }{\left\{ \mathcal{E}_{kn}^{1}\left( \xi _i \right)E_{n}^{1}+\mathcal{F}_{kn}^{1}\left( \xi _i \right)F_{n}^{1} \right\}} + \sum\limits_{n=0}^{\infty}{\left\{ \mathcal{G}_{kn}^{1}\left( \xi_i \right)G_{n}^{1}+\mathcal{H}_{kn}^{1}\left( \xi_i \right)H_{n}^{1} \right\}},
\end{align}
where $\mathcal{A}^1_{kn}$, $\mathcal{B}^1_{kn}$ etc. are integrals involving the $\rho$ and $z$ components of $\nabla c^b$. These are given by:
\begin{equation}\label{A_m_kn}
    \mathcal{A}_{kn}^{1}(\xi_i)=\int\limits_{-1}^{1}{P_{k}^{1}\left( \mu  \right)P_{n}^{1}\left( \mu  \right)\left[ \frac{\sinh \left( \lambda \xi_i  \right)}{2{{\Gamma }^{3/2}}}\mathcal{G}_{z}^{C}\left( \xi_i ,\mu  \right)+\frac{\sqrt{1-\mu^2}}{2{{\Gamma }^{3/2}}}\mathcal{G}_{\rho}^{C}\left( \xi_i ,\mu  \right) \right]\sinh \left\{ \left( n+1/2 \right)\lambda \xi_i  \right\}d\mu }.
\end{equation}

\begin{equation}\label{B_m_kn}
    \mathcal{B}_{kn}^{1}(\xi_i)=\int\limits_{-1}^{1}{P_{k}^{1}\left( \mu  \right)P_{n}^{1}\left( \mu  \right)\left[ \frac{\sinh \left( \lambda \xi_i  \right)}{2{{\Gamma }^{3/2}}}\mathcal{G}_{z}^{C}\left( \xi_i ,\mu  \right)+\frac{\sqrt{1-\mu^2}}{2{{\Gamma }^{3/2}}}\mathcal{G}_{\rho}^{C}\left( \xi_i ,\mu  \right) \right]\cosh \left\{ \left( n+1/2 \right)\lambda \xi_i  \right\}d\mu }.
\end{equation}

\begin{equation}\label{C_m_kn}
    \mathcal{C}_{kn}^{1}(\xi_i)=\int\limits_{-1}^{1}{P_{k}^{1}\left( \mu  \right)P_{n}^{1}\left( \mu  \right) \Gamma^{-1/2} \mathcal{G}_{z}^{C}\left( \xi_i ,\mu  \right) \sinh \left\{ \left( n+1/2 \right)\lambda \xi_i  \right\}d\mu }.
\end{equation}

\begin{equation}\label{E_m_kn}
    \mathcal{E}_{kn}^{1}(\xi_i)=\int\limits_{-1}^{1}{P_{k}^{1}\left( \mu  \right)P_{n}^{2}\left( \mu  \right) \frac{\Gamma^{-1/2}}{2} \mathcal{G}_{\rho}^{C} \left( \xi_i ,\mu  \right) \sinh \left\{ \left( n+1/2 \right)\lambda \xi_i  \right\}d\mu }.
\end{equation}

\begin{equation}\label{F_m_kn}
    \mathcal{F}_{kn}^{1}(\xi_i)=\int\limits_{-1}^{1}{P_{k}^{1}\left( \mu  \right)P_{n}^{2}\left( \mu  \right) \frac{\Gamma^{-1/2}}{2} \mathcal{G}_{\rho}^{C} \left( \xi_i ,\mu  \right) \cosh \left\{ \left( n+1/2 \right)\lambda \xi_i  \right\}d\mu }.
\end{equation}

\begin{equation}\label{G_m_kn}
    \mathcal{G}_{kn}^{1}(\xi_i)=\int\limits_{-1}^{1}{P_{k}^{1}\left( \mu  \right)P_{n}^{0}\left( \mu  \right) \frac{\Gamma^{-1/2}}{2} \mathcal{G}_{\rho}^{C} \left( \xi_i ,\mu  \right) \sinh \left\{ \left( n+1/2 \right)\lambda \xi_i  \right\}d\mu }.
\end{equation}

\begin{equation}\label{H_m_kn}
    \mathcal{H}_{kn}^{1}(\xi_i)=\int\limits_{-1}^{1}{P_{k}^{1}\left( \mu  \right)P_{n}^{0}\left( \mu  \right) \frac{\Gamma^{-1/2}}{2} \mathcal{G}_{\rho}^{C} \left( \xi_i ,\mu  \right) \cosh \left\{ \left( n+1/2 \right)\lambda \xi_i  \right\}d\mu }.
\end{equation}
In Eqs.~\eqref{A_m_kn} to \eqref{H_m_kn}, $\mathcal{G}_{\rho}^{C}$ (resp. $\mathcal{G}_{z}^{C}$) is the $\rho$-component (resp. $z$-component) of the gradient of the base state concentration, $\nabla c^b$. These can be evaluated by using the following expression for $\nabla c^b$,
\begin{align}\label{grad_cb}
  \nabla c^b & = \sum\limits_{n=0}^{\infty }{\left\{ \frac{{{\Gamma }^{3/2}}}{a\lambda }\frac{d c_n}{d \xi }{{L}_n}\left( \mu  \right)+\frac{{{\Gamma }^{1/2}}\sinh \left( \lambda \xi  \right)}{2a}{{L}_n}\left( \mu  \right)c_n\left( \xi  \right) \right\}{{\mathbf{e}}_{\xi }}} \nonumber\\ 
 & +\sum\limits_{n=0}^{\infty }{\left\{ \frac{{{\Gamma }^{3/2}}\sqrt{1-\mu^2}}{a}{{{{L}'}}_n}\left( \mu  \right)c_n\left( \xi  \right)-\frac{{{\Gamma }^{1/2}}\sqrt{1-\mu^2}}{2a}{{L}_n}\left( \mu  \right)c_n\left( \xi  \right) \right\}{{\mathbf{e}}_{\mu }}},
\end{align}
and, the relationship between the basis vectors in the bi-spherical and cylindrical coordinate systems, given in Eqs.~\eqref{erho_ez_exi_emu}.

\subsection{Advection of the base state solute distribution by the drop motion}

The projection of the $-\textbf{V}' \cdot \nabla c^b$ term, after evaluation at $\xi = \xi_i$, onto $P^1_k(\mu) \cos(\phi)$ is:
\begin{equation}\label{disc_U_grad_cb}
    \left. \left< -\textbf{V}' \cdot \nabla c^b \right> \right|_{\xi = \xi_i}  = \mathbf{B}^7 V'_x = B^7_k\left( \xi_i \right) V'_x,
\end{equation}
where,
\begin{equation}
    B^7_k\left( \xi_i \right) = -\int\limits_{-1}^{1}{\frac{\mathcal{G}_{\rho }^{C}\left( {{\xi }_{i}},\mu  \right)}{\Gamma} P_{k}^{1}\left( \mu  \right)d\mu }.
\end{equation}

\subsection{Solute diffusion}

The projection of the $\nabla^2 c'$ term, after evaluation at $\xi = \xi_i$, onto $P^1_k(\mu) \cos(\phi)$ is:
\begin{align}\label{disc_lap_cprime}
    \left. \left< \nabla^2 c' \right> \right|_{\xi = \xi_i} & = \mathbf{A}^3 \cdot \mathbf{C}' + \mathbf{A}^4 \cdot \frac{d^2 \mathbf{C}'}{d \xi^2 }, \nonumber \\ & =
\sum\limits_{n=1}^{\infty }{A_{kn}^3\left( \xi_i \right) c^1_n\left( {{\xi }_{i}} \right)} + \sum\limits_{n=1}^{\infty }{A_{kn}^4\left( \xi \right) \left.\frac{d^2 c^1_n}{d \xi^2}\right|_{\xi=\xi_i} },
\end{align}
where $A^3_{kn}\left( \xi_i \right)$ and $A^4_{kn}\left( \xi_i \right)$ are integrals that depend only on $\xi$, as shown below:
\begin{equation}\label{sc_D_zero_kn}
    A^3_{kn}\left( \xi_i \right) = -\frac{\sqrt{2}}{{{a}^{2}}}\sum\limits_{p=0}^{\infty }{{{\left( n+\frac{1}{2} \right)}^{2}}D_{knp}^{1}{\textrm{e}^{-\left( p+1/2 \right)\left| \lambda {{\xi }_{i}} \right|}}},
\end{equation}
and,
\begin{equation}\label{sc_D_minus_kn}
    A^4_{kn}\left( \xi_i \right) = \frac{\sqrt{2}}{{{a}^{2}}{{\lambda }^{2}}}\sum\limits_{p=0}^{\infty }{D_{knp}^{1}{\textrm{e}^{-\left( p+1/2 \right)\left| \lambda \xi_i  \right|}}}.
\end{equation}
The tensor $D^1_{knp}$ in Eqs.~\eqref{sc_D_zero_kn} and \eqref{sc_D_minus_kn} is:
\begin{equation}\label{D1_knp}
    D^1_{knp} = {{\cosh }^{2}}\left( \lambda {{\xi }_{i}} \right)T_{knp}^{0}-2\cosh \left( \lambda {{\xi }_{i}} \right)T_{knp}^{1}+T_{knp}^{2},
\end{equation}
with, $T^{i}_{knp}$ being integrals involving the appropriate combination of a zero- and first-order Legendre polynomials:
\begin{equation}\label{T_int}
    T^i_{knp} = \int\limits_{-1}^{1}{\mu^i P^1_k \left( \mu \right) P^1_n \left( \mu \right) L_p\left( \mu \right) d\mu}.
\end{equation}
In our numerical implementation, we do not pursue any analytical simplifications that utilize the properties of the associated Legendre polynomials, instead we numerically evaluate all the complicated integrals appearing above using suitable Gaussian quadratures.

\providecommand{\noopsort}[1]{}\providecommand{\singleletter}[1]{#1}%

\end{document}